\documentclass[12pt]{iopart}
\usepackage{epsfig,amssymb,psfig}

\def\spose#1{\hbox to 0pt{#1\hss}}
\def\lesssim{\mathrel{\spose{\lower 3pt\hbox{$\mathchar"218$}}
 \raise 2.0pt\hbox{$\mathchar"13C$}}}
\def\gtrsim{\mathrel{\spose{\lower 3pt\hbox{$\mathchar"218$}}
 \raise 2.0pt\hbox{$\mathchar"13E$}}}

\def\<{\langle}
\def\>{\rangle}
\newcommand\onlinecite{\cite}

\begin{document}

\title{ 
Relaxational dynamics in 3D randomly diluted Ising models
} 

\author{Martin Hasenbusch,$^{1}$
Andrea Pelissetto,$^2$
and Ettore Vicari$\,^1$ }

\address{$^1$ Dip. Fisica dell'Universit\`a di Pisa and
INFN, I-56127 Pisa, Italy}

\address{$^2$
Dip. Fisica dell'Universit\`a di Roma ``La Sapienza" and INFN, \\
I-00185 Roma, Italy}

\ead{
Martin.Hasenbusch@df.unipi.it,
Andrea.Pelissetto@roma1.infn.it,
Ettore.Vicari@df.unipi.it}

\begin{abstract}
  
  We study the purely relaxational dynamics (model A) at criticality in
  three-dimensional disordered Ising systems whose static critical behaviour
  belongs to the randomly diluted Ising universality class. We consider the
  site-diluted and bond-diluted Ising models, and the $\pm J$ Ising model
  along the paramagnetic-ferromagnetic transition line. We perform Monte Carlo
  simulations at the critical point using the Metropolis algorithm and study
  the dynamic behaviour in equilibrium at various
  values of the disorder parameter.  The results provide a robust evidence of
  the existence of a unique model-A dynamic universality class which describes
  the relaxational critical dynamics in all considered models. In particular,
  the analysis of the size-dependence of suitably defined 
  autocorrelation times at the critical point 
  provides the estimate $z=2.35(2)$ for the universal dynamic 
  critical exponent.
  We also study the off-equilibrium relaxational dynamics following a quench
  from $T=\infty$ to $T=T_c$.  In agreement with the 
  field-theory scenario, the analysis of the off-equilibrium 
  dynamic critical behavior
  gives an estimate of $z$ that is perfectly consistent with the equilibrium
  estimate $z=2.35(2)$.

\end{abstract}



\maketitle


\section{Introduction}
\label{intro}

Randomly diluted uniaxial antiferromagnets, for instance,
Fe$_{p}$Zn$_{1-p}$F${}_2$ and Mn$_{p}$Zn$_{1-p}$F${}_2$, \, have been much
investigated experimentally and theoretically
\cite{Belanger-00,PV-02,FHY-03,JBCBH-05}. For sufficiently low impurity
concentration $1-p$, these systems undergo a second-order phase transition at
$T_c(p) < T_c(p=1)$. The critical behaviour is approximately independent of
the impurity concentration and definitely different from the one of the pure
system.  These results have been successfully explained by the
field-theoretical (FT) renormalisation group (RG), which predicts the presence
of a single universality class associated with the paramagnetic-ferromagnetic
transition that occurs in Ising systems with quenched random dilution. Monte
Carlo (MC) results have been contradictory for a long time, finding
model-dependent critical exponents. In \cite{BFMMPR-98} this apparent
non-universality was shown to be an effect of strong scaling corrections.
They are slowly decaying due to the fact that the leading
correction-to-scaling exponent $\omega$ is quite small: $\omega = 0.29(2)$
(see \ref{omegaest}).

The analyses significantly gain accuracy when using improved Hamiltonians, for
which the leading scaling corrections are suppressed for any thermodynamic
quantity, and improved estimators, which are such that the leading scaling
correction is suppressed for any model in the same universality class.  MC
simulations of different improved Hamiltonians \cite{HPPV-07,HPPV-07-2}
confirmed that the static critical behaviour is model-independent, in
agreement with the FT description, and provided accurate estimates
of the static critical exponents, $\nu=0.683(2)$ and $\eta=0.036(1)$
\cite{HPPV-07,CMPV-03,BFMMPR-98}. They are in good agreement with the FT
perturbative results \cite{PV-00} $\nu=0.678(10)$ and $\eta=0.030(3)$ obtained
by the analysis of high-order (six-loop) perturbative expansions (similar
results are obtained at five loops \cite{PS-00}).  The apparent
non-universality observed in previous numerical works was mainly due to the
fact that scaling corrections were neglected. As a consequence, previous
studies did not really observe the asymptotic critical behaviour and only
determined effective exponents depending on all parameters of the investigated
model.

In this paper we extend the analysis to the critical dynamics.  We consider a
purely relaxational dynamics without conserved order parameters, also known as
model A \cite{HH-77}, as appropriate for uniaxial magnetic materials. 
Experimental
results are reported in \onlinecite{RHE-92,BFJKLM-88,Barrett-86}. According to
the FT RG (see, e.g., \onlinecite{JOS-95,CG-05,FM-06}), the dynamic behaviour
should be the same in all RDIs systems, as is the case for the static
criticality.  Moreover, the leading scaling corrections appearing in dynamical
quantities should be associated with the same RG operators that control the
nonasymptotic behaviour of static quantities and thus, they should be
characterized by the same exponents as in the static case, i.e., by
$\omega=0.29(2)$ and $\omega_2=0.82(8)$.  As a consequence, in the case of
improved Hamiltonians, leading scaling corrections should also be absent in
dynamical quantities. Therefore, the most precise estimates of dynamic
universal quantities should be obtained in improved models, as in the static
case.

Previous MC studies \cite{PV-93,Heuer-93,PRR-99,IOK-99,SP-05,IIBH-06} of
equilibrium and off-equilibrium dynamics apparently have not confirmed the FT
general predictions. They have mainly focused on the dynamic critical exponent
$z$, which characterizes the divergence of the autocorrelation times when
approaching the critical point.  In most of the cases they have found that $z$
is model dependent and have provided estimates which range from $z\approx 2.1$
to $z\approx 2.6$, depending apparently on the method, the favoured values of
the dilution parameter $p$, whether it is determined from equilibrium or
off-equilibrium simulations, etc. In \cite{PRR-99,SP-05} the universality of
$z$ was verified, obtaining $z\approx 2.6$, but the leading scaling-correction
exponent was not consistent with the static one, as predicted by the FT
approach. Moreover, this result is inconsistent with the perturbative FT
estimate obtained from analyses of the perturbative expansions
\cite{GMM-77,DeDominicis-78,PV-92,JOS-95,PBOF-98,BDFH-05,PPK-06} at two and
three loops, which suggest $z\approx 2.18$.

In this paper we study three disordered Ising systems whose static critical
behaviour belongs to the 3D RDIs universality class: the randomly site-diluted
Ising model (RSIM), the randomly bond-diluted Ising model (RBIM), and the $\pm
J$ Ising model along the paramagnetic-ferromagnetic transition line. Their
static critical behaviour was carefully investigated in
\cite{HPPV-07,HPPV-07-2}.  In particular, the value $p^*$ of the dilution
parameter corresponding to an improved model was determined for each of them.
We simulate these models by using the Metropolis algorithm (with a suitable
modification in the case of the RSIM and RBIM to avoid ergodicity problems,
see \ref{appmetro}), which does not satisfy any conservation law, and thus
allows us to investigate the model-A dynamics.  We consider cubic lattices of
size $L^3$ with $8\le L \le 64$.

The main purpose is to check whether the dynamic critical behaviour is
consistent with the FT RG, that is with the existence of a unique model-A
universality class for RDIs systems.  We focus on the dynamic critical
exponent $z$, and determine it in the RSIM, the RBIM, and the $\pm J$ Ising
model.  We find that the autocorrelation times extracted from the
autocorrelation function of the magnetic suspectibility at $T_c$ behave as
\begin{equation}
\tau = c L^{z} (1 + c_{11} L^{-\omega} + c_{12} L^{-2\omega} + \cdots +
c_{21} L^{-\omega_2} + \cdots)
\label{taufss}
\end{equation}
with a universal value of the dynamic exponent $z$. Moreover, $\omega$ and
$\omega_2$ are consistent with the static scaling-correction exponents
$\omega=0.29(2)$ and $\omega_2=0.82(8)$.  We obtain the estimates
$z=2.355(16)$, $z=2.335(18)$, and $z=2.345(17)$, respectively for the RSIM,
the RBIM, and the $\pm J$ Ising model at $p \approx p^*$. They are in good
agreement, strongly supporting universality.  Results for other values of $p$,
both larger and smaller than $p^*$, are consistent with the estimates 
of $z$ obtained at $p \approx p^*$.  We consider
\begin{equation}
z=2.35(2)
\label{veryfinalintro}
\end{equation}
as our best estimate of $z$ for the dynamic model-A universality class of RDIs
systems. These results confirm the general picture that comes out of the FT
analysis. However, from a quantitative point of view, our estimate
significantly differs from the perturbative result $z \approx 2.18$ at three
loops \cite{BDFH-05,PPK-06}.  Apparently, perturbative FT expansions 
at this order
are not able to predict accurately the exponent $z$.

The exponent $z$ can also be determined by performing off-equilibrium
simulations, since the approach to equilibrium is controlled by the same FT
model \onlinecite{JSS-89,OJ-95,CG-05}.  As a further check of our result
(\ref{veryfinalintro}), we have performed off-equilibrium MC simulations of
the RSIM at $p=0.8$, quenching $T=\infty$ configurations to $T=T_c$. The
results show that the relaxation to equilibrium is controlled by the same
dynamic exponent obtained in equilibrium simulations, i.e. $z=2.35(2)$.
Moreover, the large-time corrections are consistent with what is predicted by
the FT RG, which relates them to the static leading and next-to-leading
scaling-correction exponents $\omega=0.29(2)$ and $\omega_2=0.82(8)$. Our
results therefore confirm the FT analysis of the off-equilibrium relaxational
dynamics \cite{JSS-89,OJ-95,CG-05}.

The paper is organized as follows.  In Sec.~\ref{models} we define the
disordered Ising models that are considered in the paper.  In Sec.~\ref{twopt}
we define the quantities that are measured in the MC simulation and discuss
the FT predictions.  In Sec.~\ref{zexp} we report the finite-size scaling
(FSS) analysis of equilibrium MC simulations of the RSIM, the RBIM, and the
$\pm J$ Ising model. In Sec.~\ref{offeq} we study the off-equilibrium
relaxational critical behaviour of the RSIM, in a quench from $T=\infty$ to
$T_c$.  Finally, we draw our conclusions in Sec.~\ref{conclusions}.  In
\ref{omegaest} we refine the estimate of the leading scaling correction
exponent, obtaining $\omega=0.29(2)$.  Some details on the MC algorithm are
discussed in \ref{appmetro}.

\section{Models}
\label{models}

We consider the randomly
site-diluted Ising model (RSIM) with Hamiltonian
\begin{equation}
{\cal H}_{\rho} = - \sum_{<xy>}  \rho_x \,\rho_y \; \sigma_x \sigma_y,
\label{Hs}
\end{equation}
where the sum is extended over all nearest-neighbour sites of a simple
cubic lattice, $\sigma_x$ are Ising spin variables, and $\rho_x$
are uncorrelated quenched random variables, which are equal to 1
with probability $p$ (the spin concentration) and 0 with
probability $1-p$ (the impurity concentration).  
We also consider the randomly bond-diluted Ising model (RBIM) in
which the disorder variables are associated with links rather than with sites. 
It is defined by the Hamiltonian
\begin{equation}
{\cal H}_{j} = - \sum_{<xy>} j_{xy} \; \sigma_x \sigma_y,
\label{Hb}
\end{equation}
where the couplings $j_{xy}$ are uncorrelated quenched random variables,
which take values 0,1 with probability distribution
\begin{equation}
P(j_{xy}) = p \delta(j_{xy} - 1) + (1-p) \delta(j_{xy}). 
\label{probrbim}
\end{equation}
Note that the exchange interaction is ferromagnetic in both models.

MC simulations \cite{HPPV-07,HPPV-07-2} have provided
strong numerical evidence that the static critical behaviour of the
RSIM (for $1>p>p_{s}$, where $p_s$ is the site-percolation point, 
$p_{s}=0.3116081(13)$ on a simple cubic lattice~\cite{BFMMPR-99}) and of the
RBIM (for $1>p>p_{b}$, where $p_b$ is the bond-percolation point,
$p_{b}=0.2488126(5)$ on a simple cubic lattice~\cite{LZ-98}) belong to the same
universality class.  The most precise estimates of the static critical 
exponents have been obtained by MC simulations:
\cite{HPPV-07,CMPV-03,BFMMPR-98} $\nu=0.683(2)$ and
$\eta=0.036(1)$. These estimates are in good agreement with the 
perturbative FT results \cite{PV-00,PS-00}
$\nu=0.678(10)$ and $\eta=0.030(3)$, and with 
experiments \cite{Belanger-00,PV-02}.
Also the leading and next-to-leading correction-to-scaling
exponents have been computed. Here we shall obtain a precise
estimate of the leading exponent $\omega$, $\omega=0.29(2)$, by a combined
analysis of the data obtained in  \onlinecite{HPPV-07} 
and those obtained in the present work; see \ref{omegaest} for details.
As for the next-to-leading exponent, we quote 
the FT estimate obtained  in \onlinecite{HPPV-07},
$\omega_2=0.82(8)$.

We also consider the $\pm J$ Ising model, defined by Hamiltonian
(\ref{Hb}) with exchange interactions $j_{xy}$ which take values $\pm 1$ with
probability distribution \cite{EA-75}
\begin{equation}
P(j_{xy}) = p \delta(j_{xy} - 1) + (1-p) \delta(j_{xy} + 1). 
\label{probpmj}
\end{equation}
Unlike the RSIM and the RBIM, 
the $\pm J$ Ising model is frustrated  for any $p$.
Nonetheless, the paramagnetic-ferromagnetic transition 
line that occurs in this model for $0 < p < 1 - p_N$ and $p_N < p < 1$
also belongs to the RDIs universality class \cite{HPPV-07-2}.
Here $p_N$ is the location of the magnetic-glassy Nishimori multicritical 
point, which has been recently computed in \onlinecite{HPPV-07-3}:
$p_{N}=0.76820(4)$.

In this work we consider a relaxational dynamics without conserved order
parameters, i.e.~the so-called model A. In lattice systems this dynamics is
usually realized by using the Metropolis algorithm. In the case of the RSIM
and of the RBIM however, if a sequential updating scheme is used, the
Metropolis algorithm with the standard acceptance 
probability $P_A=$min$[1,\exp(-\beta \Delta {\cal H})]$ 
is not ergodic and thus it does not provide the correct
dynamics.
An ergodic dynamics is obtained by introducing a simple modification
which is described in \ref{appmetro}. 
In the $\pm J$ Ising model we use the
standard Metropolis algorithm with a sequential updating scheme. In this model
the specific problem we observed in the RSIM and in the RBIM is not present
(note, however, 
that, to our knowledge, a rigorous proof of ergodicity is lacking for this
updating scheme; this is also the case of the pure Ising model).

Note that the algorithm with sequential updating does not satisfy detailed
balance and hence does not strictly correspond to a reversible
dynamics.\footnote{ The Metropolis update is obtained from a single-site
  update. If $P^{z} = \{p_{xy}^{(z)}\}$ is the transition matrix for the
  update of site $z$, $P^{z}$ satisfies the detailed-balance condition $\pi_x
  p_{xy}^{(z)} = \pi_y p_{yx}^{(z)}$. However, this does not imply that the
  dynamics is reversible. Indeed, if lattice sites are updated sequentially,
  the transition matrix for a full sweep is $P^{\rm sw} = P^{z_1} P^{z_2}
  P^{z_3} \ldots P^{z_n}$, where $n$ is the number of lattice sites. $P^{\rm
    sw}$ does not satisfy the detailed-balance condition since the matrices
  $P^{z_i}$, $P^{z_j}$ for nearest neighbours $z_i$ and $z_j$ do not commute.
  For a more detailed discussion, see, e.g., \cite{Sokal-97}. } Detailed
balance is satisfied only if the spins are updated in random order. It is
commonly accepted that these two dynamics belong to the same universality
class: these violations of detailed balance are irrelevant in the critical
limit.

\section{Autocorrelation times: definitions and critical properties}
\label{twopt}

We consider the two-point correlation function
\begin{eqnarray}
G(x_2-x_1,t_2-t_1) \equiv 
\overline{\langle \sigma(x_1,t_1) \,\sigma(x_2,t_2) \rangle} ,
\label{twop}
\end{eqnarray}
where the overline indicates the quenched average over the disorder
probability distribution and $\langle \cdot\cdot\cdot \rangle$ indicates the
thermal average.
Near the critical point correlations develop both in space and time. They can
be characterized in terms of the equal-time 
second-moment correlation length $\xi$ and of an 
autocorrelation time $\tau$. In the infinite-volume limit 
the correlation length $\xi$ can be defined as
\begin{equation}
\xi^2 \equiv 
    - {1\over \chi} \left. {\partial \widetilde{G}(k,0) \over \partial k^2} 
      \right|_{k^2 = 0} , 
\label{xidefc}
\end{equation}
where $\widetilde{G}(k,t)$ is the Fourier transform of $G(x,t)$ with respect
to the $x$ variable and
\begin{equation}
\chi\equiv\sum_x G(x,0) = \widetilde{G}(0,0)
\label{chidef}
\end{equation}
is the static magnetic susceptibility.  On a finite lattice 
with periodic boundary conditions, we define $\xi$ as
\begin{equation}
\xi^2 \equiv {\widetilde{G}(0,0) - \widetilde{G}(q_{\rm min},0) \over 
          \hat{q}_{\rm min}^2 \widetilde{G}(q_{\rm min},0) },
\label{xidef}
\end{equation}
where $q_{\rm min} \equiv (2\pi/L,0,0)$, $\hat{q} \equiv 2 \sin q/2$.  
To define the autocorrelation time, we consider the autocorrelation function
$A(t)$ of a long-distance quantity.
Then, we define the integrated autocorrelation time 
\begin{equation}
\tau_{{\rm int}} \equiv {1\over2} 
 \sum_{t=-\infty}^\infty \, {A(t)\over A(0)} = 
  {1\over 2} + \sum_{t=1}^\infty \, {A(t)\over A(0)}.
\label{tauintdef}
\end{equation}
Here $t$ is the Metropolis time and one time unit corresponds to a complete 
lattice sweep.

In the critical limit $\xi$ and the autocorrelation time $\tau_{{\rm int}}$ 
diverge.
If $t_r \equiv (T - T_c)/T_c$ and $T_c$ is the critical temperature,
for $|t_r|\to 0$ we have in the thermodynamic limit
\begin{equation}
\xi\sim |t_r|^{-\nu},
\qquad
\tau_{{\rm int}} \sim |t_r|^{-z\nu} \sim \xi^z,
\label{diver}
\end{equation}
where $\nu$ is the usual static exponent and $z$ is a dynamic exponent that
depends on the considered dynamics. 

The correlation function $\widetilde{G}(k,t)$ is the quantity of direct 
experimental interest and thus we could take $A(t) = \widetilde{G}(k,t)$. 
However, for the determination of the dynamic critical exponent $z$,
it is computationally more convenient to use a different quantity.
We consider the autocorrelation function of the magnetic susceptibility
\begin{equation}
A(t) = \overline{
\langle  S(0) S(t) \rangle
- \langle  S \rangle^2}, \qquad 
S(t) \equiv {1\over V} \Bigl[ \sum_x \sigma(x,t) \Bigr]^2.
\label{scorr}
\end{equation}
Using (\ref{tauintdef}) we could  determine
the autocorrelation time $\tau_{{\rm int}}$ and then, we could use
it to determine $z$. However, the determination of this quantity requires 
the knowledge 
of the large-$t$ behaviour of $A(t)$. Since it is difficult to determine it 
precisely, $\tau_{{\rm int}}$ is unsuitable for a high-precision study. 
We now introduce a new time scale which is particularly convenient 
numerically.
Let us define
\begin{equation}
\tau_{\rm eff}(t+n/2) \equiv {n\over \ln [A(t)/A(t+n)]},
\label{defBL}
\end{equation}
where $n$ is a fixed integer number. A linear interpolation can be used
to extend $\tau_{\rm eff}(t)$ to all real numbers.
Then, for any positive $x$, 
we define an autocorrelation time $\tau_x$ as the solution 
of the equation
\begin{equation} 
\tau_x = \tau_{\rm eff}(x\tau_x).
\label{tauxdef}
\end{equation}
This definition is based on the idea that, 
if $A(t)$ were a pure exponential, i.e.,
$A(t) = A_0 \exp(-t/\tau)$, then
$\tau_{\rm eff}(t) = \tau$ for all $t$ and
thus $\tau_x = \tau$ for any $x$.

Let us now consider the thermodynamic limit with $T > T_c$ 
(high-temperature phase) and 
let us prove that, if the autocorrelation functions 
decay faster than any power of $t$ in the critical limit,
then $\tau_x$ behaves as $\xi^z$ 
as any ``good" autocorrelation time. More precisely, we show that 
$\tau_x/\tau_{{\rm int}}$ is finite and nonzero 
in the critical limit for any finite $x$.
Since $A(t)$ is an autocorrelation function of a long-range quantity,
close to the critical point it obeys the scaling law
\begin{equation}
{A(t)\over A(0)} = f(S), \qquad\qquad  S\equiv t/\tau_{{\rm int}}.
\label{scal-A}
\end{equation}
In the critical limit and for fixed $n$, we have 
$n/\tau_{{\rm int}} \to 0$. Thus, we can expand
\begin{equation}
\tau_{\rm eff}(t+n/2) = - \tau_{{\rm int}} \times {f(S)\over f'(S)} 
    \left[1 + O(n/\tau_{{\rm int}})\right].
\label{BLexp}
\end{equation}
If we now define $a_x \equiv x \tau_x/\tau_{{\rm int}}$, we obtain
in the critical limit the equation
\begin{equation}
a_x = - x f(a_x)/f'(a_x).
\label{eqax}
\end{equation}
It is a simple matter to show that, if $f(S)$ decays faster than
any power of $S$ ($S^q f(S)\to 0$ for $S\to\infty$ and any $q$), 
there is always (at least)
one strictly positive solution $a_x$ of (\ref{eqax}).\footnote{
Proof. The function $f(y)$ is expected to be positive and
strictly decreasing, so that 
$f(y) > 0$ and $f'(y) < 0$ for any $y$. 
Since $y^q f(y) \to 0$ for $y\to\infty$ and $f(y) > 0$,
$y^q f(y)$ decreases for large values of $y$. Therefore,
we have $(y^q f(y))' = q y^{q-1} f(y) + y^q f'(y) < 0$. 
This implies $y f'(y)/f(y) < - q$ for $y$ large enough. 
Since $q$ can be arbitrarily large, this implies 
$y f'(y)/f(y) \to -\infty$ for $y\to \infty$. To end the proof,
define $h(y) = y + x f(y)/f'(y)$. For
$y = 0$ we have $h(0) = x f(0)/f'(0) < 0$. For
$y\to\infty$, we have
$h(y) = y [1 + x f(y)/(y f'(y))] \approx y \to +\infty$
(here we use the result $y f'(y)/f(y)\to -\infty$ for $y\to\infty$).
The function $h(y)$ is therefore negative for small
$y$ and positive for large $y$. Since it is continuous, $h(y)$ must vanish
at a finite nonvanishing value of $y$.}
Thus, we have proved that, for any $x>0$, the ratio
$\tau_x/\tau_{{\rm int}}$ is finite and strictly positive 
in the critical limit. It follows that 
$\tau_x$ diverges as $\xi^z$ in the critical limit. 

The condition that $f(S)$ decays faster than any power of $S$ is obviously
satisfied if
$f(S)$ decays exponentially, i.e. if $f(S) \approx A S^a \exp(- bS)$ 
for large $S$, where $a$ is some exponent.
While an exponential decay
of the correlations is obvious in pure ferromagnetic systems for 
temperatures $T > T_c$, in the case of random systems some discussion is 
needed.  Indeed, in dilute systems one expects a non-exponential relaxation 
for large values of $t$ \cite{Bray-88}, due to the presence of rare
compact clusters without vacancies that are fully magnetized at temperatures 
that are below the critical temperature of the pure system (the same 
clusters are responsible for the weak Griffiths singularities in 
the high-temperature free energy \cite{Griffiths}). For instance,
in Ising systems the infinite-volume 
spin-spin autocorrelation function $G(x=0,t)$ 
is expected to decay as \cite{Bray-88,DRS-88,Bray-89,CMM-98} 
\begin{equation}
G(x=0,t) \approx B \exp [- C (\ln t)^{3/2} ]
\label{G-non-exp}
\end{equation}
for $t\to \infty$. In the infinite-volume limit 
also $A(t)$ may show a non-exponential 
behavior for large $t$ in the high-temperature phase. However, note 
that this does not necessarily imply that the scaling function 
defined in (\ref{scal-A})
decays non-exponentially. On the contrary, one can argue \cite{Bray-88}
that the Griffiths tail (\ref{G-non-exp})
becomes irrelevant in the critical limit.
This is essentially due to the fact that $B$ and $C$ 
that appear in (\ref{G-non-exp}) are expected to be smooth functions of
the temperature that approach finite constants as 
$T\to T_c$. Thus, in the critical limit, $t\to\infty$, $T\to T_c$
at fixed $S$, the non-analytic contribution simply vanishes.\footnote{
This phenomenon can be easily 
understood if one imagines $A(t)$ to have the 
form $A_1 \xi^a \exp(-B_1 t \xi^{-z}) + A_2 \exp (- B_2 (\ln t)^{3/2})$.
The first term is the critical contribution, while the second one is the 
non-exponential Griffiths tail. 
The second term dominates for 
$t \gg t^*$, where $t^*$ is the value of $t$ at which 
the two terms have the same magnitude.
In the critical limit
we have $t^*\sim \xi^z (\ln \xi)^{3/2}$. Since the critical limit is taken
at $t/\xi^{z}$ fixed, the relevant quantity is $t^*/\xi^{z}$, which diverges 
as $(\ln \xi)^{3/2}$, as $T\to T_c$. This means that, for any fixed value
of $S \equiv t/\tau_{\rm int}\sim t/\xi^{z}$, 
sufficiently close to the critical temperature, $t$ always 
satisfies the condition $t\ll t^*$, i.e. belongs to the region in which 
the non-exponential tail (\ref{G-non-exp}) is negligible. 
These considerations also
indicate that one should limit oneself to times $t \ll t^*$ in studies of the 
infinite-volume critical behavior in the 
high-temperature phase. Therefore, one should always choose $x$ so that 
$\tau_x \ll t^*$ for all considered systems. Otherwise, the extrapolated
critical behavior would be incorrect.}

In the above-presented discussion, $\tau_x$ represents an infinite-volume
autocorrelation time determined in the high-temperature phase. 
A similar discussion applies if we consider the FSS behavior. For instance,
at $T_c$ we have
\begin{equation}
{A(t,L)\over A(0,L)} = f_{FSS}(\widehat{S}), \qquad\qquad  
      \widehat{S}\equiv t L^{-z},
\label{scal-A-FSS}
\end{equation}
where $L$ is the lattice size. The function $A(t,L)$ decays exponentially
for any $L$ (this is rigorously true for an aperiodic dynamics in 
a discrete spin system). This fact does not necessarily imply that 
$f_{FSS}(\widehat{S})$ decays exponentially 
(a non-exponential behavior could occur if the 
exponential decay sets in for $t > t^*\sim L^{z+\epsilon}$, $\epsilon > 0$),
though the discussion presented above makes this possibility quite unlikely.
In any case, if $f_{FSS}(\widehat{S})$ decays faster than any power 
of $\widehat{S}$,
the previous proof indicates that $\tau_x/L^z$ is finite in the critical limit
for any finite $x$, and thus $\tau_x$ is a good autocorrelation time.

Beside the integrated autocorrelation time one can also define 
an exponential autocorrelation time:
\begin{equation}
\tau_{{\rm exp}} \equiv - \lim_{|t|\to\infty}
      {|t|\over \ln A(t)}\; .
\label{tauedef}
\end{equation}
This quantity is well defined in a finite volume since $A(t,L)$ decays 
exponentially, but, as a consequence of (\ref{G-non-exp}), 
it diverges in the infinite-volume limit for all 
$T_c \le T \le T_c(p=1)$.
As a consequence, in the infinite-volume limit at fixed temperature, 
$\tau_x$ diverges as $x\to \infty$.
However, the decoupling of the non-exponential tail in the critical limit
implies that 
\begin{equation}
\lim_{x\to\infty} \lim_{T\to T_c+} {\tau_x\over \xi^z} 
\label{tauexp-lim}
\end{equation}
is finite and related to the decay rate of $f(S)$ for large $S$. 
Of course, the two limits in (\ref{tauexp-lim}) cannot be interchanged.

On a finite lattice of size $L$, $\tau_{{\rm exp}}$ is always well defined. 
Nonetheless, this 
does not imply that $\tau_{{\rm exp}}$ is a good autocorrelation time.
On the contrary, at $T=T_c$ we  expect 
$\tau_{{\rm exp}}/L^z$ to diverge as $L\to \infty$. 
Indeed, for each $L$, $\tau_{{\rm exp}}(L)$ is always given by the 
decay rate of the autocorrelation function for the slowest sample, 
however small is the amplitude 
of this contribution to the autocorrelation function 
(note that, for finite values of $L$, the disorder average is 
a finite sum). As a consequence,
$\tau_{{\rm exp}}(L)$ is the exponential autocorrelation time for 
a pure Ising system in the low-temperature phase, which is expected to 
increase faster than any power of $L$, as $L\to\infty$ [if tunnelling 
events dominate $\tau_{\rm exp}(L)\sim \exp(2\sigma L^2)]$.  
Therefore, $\tau_{\rm exp}/L^z\to\infty$ as $L\to \infty$. 
The irrelevance of the Griffiths phenomenon in the critical limit 
should however imply that 
\begin{equation}
 \lim_{x\to\infty} \lim_{L\to\infty} {\tau_x(T_c,L)\over L^z}
\end{equation}
is finite and related to the decay rate of $f_{FSS}(\widehat{S})$. 
This is the finite-volume analogue of (\ref{tauexp-lim}).

In the definition (\ref{defBL}) the integer $n$ can be taken arbitrarily.
However, the asymptotic critical behaviour is observed only if $n \ll
\tau_{{\rm int}}$, see (\ref{BLexp}). Therefore, in practice $n$ should not
be too large. It is also convenient to take $n$ not too small, since this
avoids computing $A(t)$ too frequently in the MC simulations.  Note also that,
when $n$ decreases, the errors on $\tau_{\rm eff}(t)$ increase since $A(t)$
and $A(t+n)$ are close. The effect is however small, because of statistical
correlations that also increase as $n$ decreases.  In our work we have always
considered values of $n$ much smaller than $\tau_x$ (typically $n \lesssim
\tau_x/20$) and we have verified that the estimate of the autocorrelation
times are independent of the chosen (small) value of $n$.

Definition (\ref{defBL}) provides an effective exponential autocorrelation
time at a finite time scale. In the same spirit, one can also define 
truncated integrated autocorrelation times. Define 
\begin{equation}
I(k) \equiv  {1\over 2} + {1\over A(0)} \sum_{t=1}^k A(t)
\end{equation}
for any integer $k$, and $I(t)$ for any real $t$ by linear interpolation.
Then, we can define 
an autocorrelation time $\tau_{x,\rm int}$ as the solution of the equation
\begin{equation}
\tau_{x,\rm int} = I(x\tau_{x,\rm int}).
\label{tauxint-def}
\end{equation}
For any $x$, this definition provides a good autocorrelation time,
which converges to $\tau_{{\rm int}}$ for $x\to \infty$. This definition
is similar to that proposed in \onlinecite{MS-88};  note, however, 
the completely different spirit in the two definitions. 
In \onlinecite{MS-88} the method was proposed 
as a practical self-consistent method for the determination of 
$\tau_{{\rm int}}$ and for this reason $x$ had to be large (in practice
$x$ was usually taken between 5 and 10). Instead, 
if one is not interested in determining $\tau_{{\rm int}}$ but only in 
computing $z$, $x$ can be taken at will. 

In this paper we compute the exponent $z$ from 
the volume dependence of an autocorrelation time at the critical temperature.
Including scaling corrections, we expect a behaviour 
of the form
\begin{equation}
\tau = c L^z \, \left(1 + c_{11} L^{-\omega} + c_{12} L^{-2\omega} + \cdots +
c_{21} L^{-\omega_2} + \cdots \right),
\label{taufss2}
\end{equation}
where $\omega$ and $\omega_2$ are the leading and next-to-leading 
critical exponents.
As in \onlinecite{HPPV-07,HPPV-07-2} we also consider the dynamical
behaviour at a fixed value of a renormalized coupling constant. Also
in this case autocorrelation times behave as in (\ref{taufss2}).

In order to determine $z$ it is crucial to have some knowledge of the
correction-to-scaling exponents that appear in (\ref{taufss2}).  RG predicts
that the static correction-to-scaling exponents also occur in dynamic
quantities. For instance, if $\chi\equiv \widetilde{G}(0,0)$ behaves as $a
L^{2-\eta} (1 + e_{11} L^{-\omega_{\rm stat}})$ at criticality for
$L\to\infty$, then a correction term decaying as $L^{-\omega_{\rm stat}}$ is
also expected in $\widetilde{G}(0,t)$ for any $t\not=0$. However, dynamics
gives also rise to new scaling corrections and they may decay slower than the
static ones (for instance, this occurs in the model-C dynamics, see
Sec.~\ref{conclusions}). In this paper we make the assumption that no new
scaling corrections with exponent less than $\omega_2 = 0.82(8)$ appear, as
indicated by the FT description of the model-A dynamics. As we shall see, this
will be confirmed by our numerical analysis.  Thus, in (\ref{taufss2})
$\omega$ and $\omega_2$ should be identified with the static
scaling-correction exponents.

In our analysis, we make use of improved models, which are such that
the leading scaling correction with exponent $\omega$ vanishes. 
Since ratios of leading scaling-correction amplitudes 
are universal (both in 
static and in dynamic correlation functions), this cancellation also occurs
in dynamic quantities. 
Improved models have been determined in \onlinecite{HPPV-07,HPPV-07-2}:
the RSIM at $p^*=0.800(5)$, the RBIM at $p^*=0.54(2)$, and the 
$\pm J$ Ising model at $p^*=0.883(3)$ are improved.
In these models the scaling
corrections proportional to $L^{-k\omega}$
vanish, so that the leading correction-to-scaling exponent is $\omega_2$. 
Therefore, numerical studies of improved models are expected to provide the
most precise estimates of universal quantities.  Of course, this
is true only if the usual model-A FT description is correct; 
otherwise, there could 
be corrections with a new dynamic exponent $\omega_{\rm dyn} < \omega_2$, 
which do not cancel and may give rise to large corrections even in models
that are improved for static quantities. 
A stringent check of this picture should be the fact that
the three different improved models we consider give consistent
results.

\section{Equilibrium estimate of the dynamic critical exponent $z$}
\label{zexp}

\subsection{Monte Carlo simulations}
\label{MC}

We perform MC simulations of the RSIM, the RBIM, and the $\pm J$ Ising
models for various values of $p$, close to the critical temperature
on cubic lattices of size $L^3$ with $L\le 64$ and 
periodic boundary conditions.
We use the Metropolis algorithm with multispin coding as 
described in \ref{appmetro}.
 
\begin{table}
\caption{
MC estimates of $\tau_x(L)$ for the RSIM at
$p=0.8$ and for various values of $x$ at $\beta=0.285744$. 
For $x=1$ we also report estimates of $\tau_{x}(L)$ extrapolated to
$\beta_c = 0.2857431(3)$.
We also report the value of $n$ that enters in the definition (\ref{defBL}).
}
\label{tabrsimp0.8}
\begin{center}
\begin{tabular}{rrlllll}
\hline
\multicolumn{1}{c}{$L$}&
\multicolumn{1}{c}{$n$}&
\multicolumn{1}{c}{$x=0.6$}&
\multicolumn{1}{c}{$x=1$}&
\multicolumn{1}{c}{$x=1$ at $\beta_c$}&
\multicolumn{1}{c}{$x=1.5$}&
\multicolumn{1}{c}{$x=2$}
\\ \hline
8 & 1 & 7.311(5) & 7.946(8) & 7.946(8) & 8.342(15) & 8.535(25) \\
12 & 2 & 18.016(10) & 19.827(17) & 19.826(17) & 20.88(3) & 21.35(5) \\
16 & 2 & 34.783(20) & 38.42(3) & 38.42(3) & 40.57(6) & 41.57(10) \\
24 & 4 & 88.47(5) & 98.21(8) & 98.20(8) & 103.78(14) & 106.51(24) \\
32 & 6 & 172.25(9) & 191.64(16) & 191.61(16) & 202.8(3) & 207.9(5) \\
48 & 16 & 442.4(3) & 494.0(6) & 493.8(6) & 523.1(1.0) & 538.5(1.7) \\
64 & 30 & 864.0(1.1) & 966.4(2.0) & 966.0(2.0) & 1024(3) & 1052(7) \\
\hline
\end{tabular}
\end{center}
\end{table}

\begin{table}
\caption{
MC estimates of $\tau_{x=1}(L)$ for the RSIM at $p=0.85$ and
$\beta_c=0.2661561(5)$, 
$p=0.8$ and $\beta_c=0.2857431(3)$, and 
$p=0.65$ and $\beta_c=0.370168(2)$, and for the 
RBIM at $p=0.7$ and $\beta_c = 0.326710(3)$, and
$p=0.55$ and $\beta_c=0.432291(2)$.
}
\label{tabtaux1}
\begin{center}
\begin{tabular}{rlllll}
\hline
\multicolumn{1}{c}{$L$}&
\multicolumn{1}{c}{RSIM $p=0.85$}&
\multicolumn{1}{c}{RSIM $p=0.8$}&
\multicolumn{1}{c}{RSIM $p=0.65$} &
\multicolumn{1}{c}{RBIM $p=0.7$}&
\multicolumn{1}{c}{RBIM $p=0.55$}
\\ \hline
8  & 7.595(7)   & 7.946(9) & 10.343(10) & 9.410(19) & 12.853(14)  \\ 
12 & 18.322(13) & 19.826(17) & 30.79(3) & 22.746(22)& 33.30(3) \\ 
16 & 34.564(24) & 38.42(3) & 67.55(6)   & 42.64(4) &  65.41(5)    \\
24 & 84.94(5)   & 98.20(8) & 204.66(22) & 103.45(6) & 169.03(11) \\
32 & 161.15(8) & 191.61(16) & 447.7(7)  & 193.56(10) & 331.38(21) \\
48 & 398.0(4)  & 493.8(6)  & 1326(3)  & 468.5(5) &  853.2(1.0)   \\
64 & 756.6(1.4) & 966.0(2.0) & 2846(12) & 874.5(1.9) & 1676(5) \\
\hline
\end{tabular}
\end{center}
\end{table}

\begin{table}
\caption{
MC estimates of $\tau_{x=1}(L)$ for the $\pm J$ Ising model
at various values of $p$ and 
at fixed $\xi/L=0.5943$.
}
\label{tabtaux1bf}
\begin{center}
\begin{tabular}{rlll}
\hline
\multicolumn{1}{c}{$L$}&
\multicolumn{1}{c}{$\pm J$ Is $p=0.83$}&
\multicolumn{1}{c}{$\pm J$ Is $p=0.883$}&
\multicolumn{1}{c}{$\pm J$ Is $p=0.9$}\\
\hline
8  & 10.260(22) & 6.507(3) & 6.035(8)   \\ 
10            & &  10.918(5) &  \\ 
12 & 29.73(5) & 16.590(7) & 14.882(10)  \\ 
14            & & 23.705(9)   & \\
16 & 63.94(11) & 32.300(17) & 28.471(16) \\
18            & & 42.353(23) &  \\
20            & & 54.13(3) &   \\
22            & & 67.47(4) &   \\
24 & 188.7(3) & 82.53(8) & 70.84(6)  \\
28            & & 118.02(17)  & \\
32 & 407.1(8) & 161.50(20) & 135.37(19) \\
40            & & 271.1(4) &  \\
48            & & 413.5(7) & 336.9(6)  \\
56            & & 592.9(1.6) & \\
64 & 2540(15) & 813(3) & 652(3) \\ 
\hline
\end{tabular}
\end{center}
\end{table}

For each lattice size we consider $N_s$ disorder samples, with $N_s$
decreasing with increasing $L$, from $N_s \approx 64\times 10^5$ for $L=8$ to
$N_s \approx 64\times 10^4$ for the largest lattice $L=64$.  Note that these
numbers of samples are much larger than those typically considered in previous
numerical studies. For each disorder sample, we
thermalize the system by using a mixture of Metropolis and
Wolff cluster updates in the case of the
RSIM and of the RBIM, while in the case of the $\pm J$ Ising model
we only used the Metropolis algorithm.\footnote{The presence of rare
disorder instances characterized by large compact clusters with no 
vacancies---those that give rise to the Griffiths tail---might be a
serious problem for the thermalization 
if only the Metropolis algorithm is used. 
If a fixed thermalization
schedule (independent of the disorder sample) is employed, the
system may be thermalized {\em on average}, but in a few rare cases the
sampling may begin much before the equilibrium state has been reached.
However, the considerations presented in
  Sec.~\ref{twopt} indicate that these contributions are irrelevant for the
  critical behavior. Moreover, their probability is quite low.  For instance,
  in the RSIM at $p=0.8$, for $L=64$ the probability of a cube of size $l=5$
  ($l=6$) without vacancies is of order $10^{-7}$ ($10^{-16}$), which should
  be compared with $1/N_s \approx 1.6\cdot 10^{-6}$.} Then, at
equilibrium, we perform runs of approximately $20\tau$ Metropolis sweeps,
where $\tau$ is the typical autocorrelation time.  The averages over disorder
are affected by a bias due to the finite number of measures at fixed disorder
\cite{BFMMPR-98-b,HPPV-07}. A bias correction is required whenever one
considers the disorder average of combinations of thermal averages.  We use
simple generalizations of the formulas reported in App.~B of
\onlinecite{HPPV-07}.\footnote{ In App.~B of \onlinecite{HPPV-07} we discuss
  the case of uncorrelated data. In our case correlations are relevant and
  thus we must somehow modify those expressions.  For instance, in order to
  compute $\overline{\langle B \rangle^2 }$, we use
\[
\overline{\langle B \rangle^2 } = 
   {4\over N_s (N_m - 2 k)^2} \sum_{\alpha=1}^{N_s}  
     \sum_{i=1}^{N_m/2 - k} B_{\alpha,i}
     \sum_{j=N_m/2 + k}^{N_m} B_{\alpha,j}
\]
where $B$ is a generic observable, $B_{\alpha,i}$ are the corresponding
MC estimates,
$N_s$ is the number of samples, $N_m$ is the number of measures in
equilibrium for each sample, and $k$ a suitable number. We have 
usually taken $k\approx 2\tau$. 
The bias is of the order $\tau^2 C_{BB}(2k)/N_m^2$, where $\tau$ 
is the integrated autocorrelation time of the variable $B$,
and $C_{BB}(t)$ the corresponding autocorrelation function.
Similar expressions are used in other cases.
}
Errors are
computed from the sample-to-sample fluctuations and are determined by using
the jackknife method.

We considered the RSIM at $p=0.8,0.65$ (which are the same values considered
in \onlinecite{HPPV-07}) and also at $p=0.85$. For the RBIM we worked at
$p=0.7,0.55$.  These runs provided new data for the static quantities that
were merged with the old ones \cite{HPPV-07} and with the results obtained 
in some additional cluster MC simulations at the largest
lattices.  They allowed us to obtain a new estimate of $\omega$ (see
\ref{omegaest}) and new estimates of $\beta_c$.  Repeating the analysis
presented in \onlinecite{HPPV-07} we obtain $\beta_c=0.2857431(3)$,
$\beta_c=0.370168(2)$, and $\beta_c=0.2661561(5)$ for the RSIM at
$p=0.8,0.65,0.85$, respectively, $\beta_c=0.432291(2)$ and $\beta_c =
0.326710(3)$ for the RBIM at $p=0.55,0.7$, respectively. For each $p$ and
$\beta$ we usually considered two values of $\beta$ very close to $\beta_c$
and determined the autocorrelation times at $\beta_c$ by linear interpolation.
For the RSIM at $p = 0.8$, runs were performed at $\beta = 0.2857440$ and
subsequently extrapolated at $\beta_c=0.2857431$ (see below).  For the $\pm J$
model we did not perform additional simulations and used the results of
\onlinecite{HPPV-07-2}. They allowed us to determine $\tau_x(L)$ at
$\xi/L=0.5943$.  In the case of the RSIM we also determined $\tau_x(L)$ at
fixed $\xi/L=0.5943$.  The results are very similar to those obtained at
$T_c$, and therefore we do not consider them in the following.

Estimates of $\tau_x(L)$ for the RSIM at $p=0.8$ are reported in
Table~\ref{tabrsimp0.8}. In the table we report the data at $\beta =
0.2857440$ and, for $x = 1$, also the extrapolations at $\beta_c$. Note that
the correction due to the small change in $\beta$ is significantly smaller
than the statistical error. 
Estimates of $\tau_x(L)$ for $x=1$ at $T_c$ for all models
and several values of $p$ are reported in Tables~\ref{tabtaux1} and
\ref{tabtaux1bf}.

\subsection{Results for the RSIM}
\label{resultsrsim}

\begin{figure*}[tb]
\centerline{\psfig{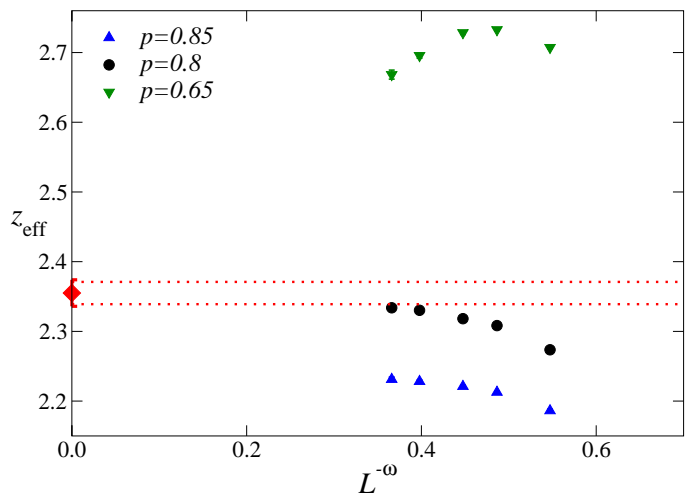}}
\vspace{2mm}
\caption{The effective exponent
$z_{\rm eff}(L)$ vs $L^{-\omega}$ with $\omega=0.29$
for the RSIM at $p=0.85,0.8,0.65$,
as obtained from $\tau_{x=1}(L)$.
The dotted lines correspond to the estimate $z=2.355(16)$ obtained 
by using the RSIM data at $p=0.8$.
}
\label{figzeffrs}
\end{figure*}

In order to determine $z$, we
define an effective exponent 
\begin{equation}
z_{\rm eff}(L) \equiv {\ln[\tau_x(2L)/\tau_x(L)]\over \ln 2},
\label{zeff}
\end{equation}
which, for $T=T_c$ and $L\to\infty$, behaves as
\begin{equation}
z_{\rm eff}(L)= z + e_{11} L^{-\omega} + e_{12} L^{-2\omega} + \cdots +
e_{21} L^{-\omega_2} + \cdots ,
\label{zefffss}
\end{equation}
see (\ref{taufss2}).
In Fig.~\ref{figzeffrs} we show $z_{\rm eff}(L)$ as obtained from the
estimates of $\tau_{x=1}(L)$ for the RSIM at $p=0.85,0.8,0.65$, reported in
Table~\ref{tabtaux1}.  The raw data show significant scaling corrections
and it is far from clear that their limit for $L\to \infty$ 
is independent of $p$.

In the following we present a detailed analysis of the MC data for 
the RSIM. First, we analyse the data
at $p=0.8$. If the FT description is correct, we should observe
a fast convergence to the infinite-volume limit, with corrections
proportional to $L^{-\omega_2}$. The results presented in 
Sec.~\ref{resultsrsimp0.8} confirm this prediction.
In particular, there is no evidence of a  
correction-to-scaling exponent smaller than $\omega_2$ 
in dynamic quantities. These results
support the general FT scenario which predicts that 
the two leading correction-to-scaling
exponents are the static ones $\omega$ and $\omega_2$. 
Then, we assume the FT scenario and 
perform a consistency check, verifying that the large differences
observed in Fig.~\ref{figzeffrs} can be explained by scaling corrections. 
To make the check more quantitative, we introduce an improved estimator
for the exponent $z$ (we use here the same strategy employed in 
\onlinecite{HPPV-07} for static quantities) and show that 
it converges to the same value obtained for $p=0.8$ with the 
expected scaling corrections. 
This allows us to confirm universality, i.e. the $p$ independence of the 
dynamic critical behaviour.

\subsubsection{Analysis at $p=0.8$.}
\label{resultsrsimp0.8}

\begin{figure*}[tb]
\centerline{\psfig{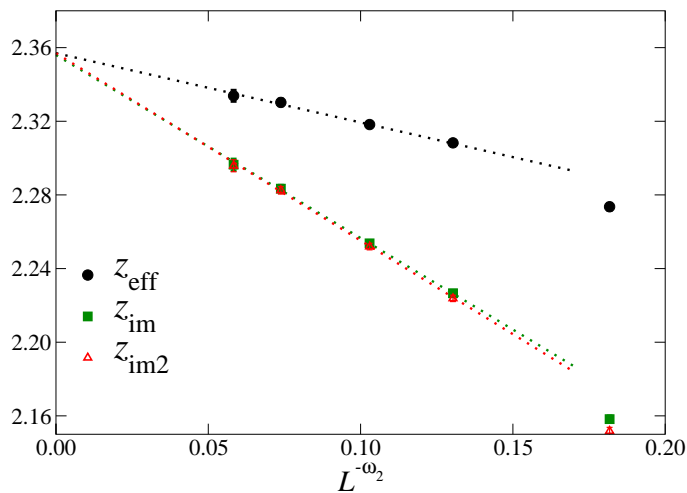}}
\vspace{2mm}
\caption{The effective exponent 
$z_{\rm eff}(L)$ and the corresponding improved quantities $z_{\rm
im}(L)$ and $z_{\rm im2} (L)$ defined in (\ref{zimdef}) 
versus $L^{-\omega_2}$ with $\omega_2=0.82$.
Results for the RSIM at $p=0.8$.
The effective exponents are obtained 
by using $\tau_x(L)$ with $x = 1$. The dotted lines 
correspond to the results of fits to $z + a L^{-0.82}$ with
$L_{\rm min}=12$.  }
\label{zeffis}
\end{figure*}

\begin{figure*}[tb]
\centerline{\psfig{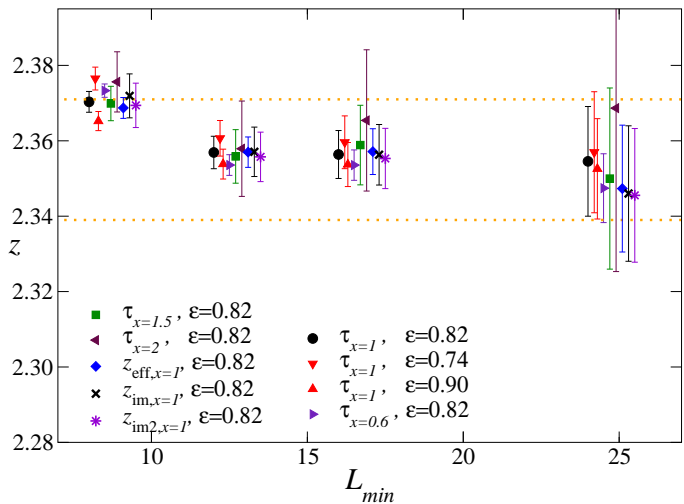}}
\vspace{2mm}
\caption{
Estimates of the dynamic exponent $z$ for the RSIM at $p=0.8$ 
obtained from fits of $\tau_x(L)$ to (\ref{acfit}) and of 
the effective exponents to $z + a L^{-\varepsilon}$. We report results
corresponding to different values of $L_{\rm min}$:
$L_{\rm min}=8,12,16,24$ (some data are slightly shifted along
the $x$ axis to make them visible).  The dotted lines correspond to the 
estimate (\ref{finalestzim}), $z=2.355(16)$.
}
\label{tausp0.8}
\end{figure*}

Let us first analyse $\tau_x(L)$ for the RSIM at $p=0.8$. If the standard FT
description of the model-A dynamics holds, the static correction-to-scaling
exponents are the most relevant ones. Since the RSIM at $p=0.8$ is improved,
the $O(L^{-k\omega})$ scaling corrections are suppressed and therefore we
expect the dominant scaling corrections to be proportional to $L^{-\omega_2}$
with $\omega_2 = 0.82(8)$.  In Fig.~\ref{zeffis} we plot $z_{\rm eff}(L)$ as
obtained from $\tau_{x=1}(L)$ versus $L^{-0.82}$.  The data with $L\ge 12$
clearly fall on a line.  To determine $z$ we assume $\tau_x(L)$ to behave as
\begin{equation}
\tau_x(L)= c L^z \left( 1 + c_{2} L^{-\varepsilon}\right)
\label{tauxfit}
\end{equation}
for $L\to \infty$, and perform fits of the form
\begin{equation}
\ln \tau_x(L) = z \ln L + \ln c + c_{2} e^{-\varepsilon \ln L},
\label{acfit}
\end{equation} 
with $\varepsilon = 0.82,0.74,0.90$, which correspond to $\omega_2 = 0.82(8)$.
Results for $x=0.6,1,1.5,2$ are shown in Fig.~\ref{tausp0.8} versus
$L_{\rm min}$, the smallest lattice size used in the fit.
They are independent of $L_{\rm min}$ for $L_{\rm min}\ge  12$, with
$\chi^2/{\rm DOF}\lesssim 1$ (DOF is the number of degrees of freedom of the 
fit).  For example, for $x=1$
and $\varepsilon=0.82$, we obtain $z=2.357(4)$ and
$c=0.0525(10)$ for $L_{\rm min}=12$, and $z=2.356(6)$ and $c=0.0526(15)$ for
$L_{\rm min}=16$.  For $x=0.6,1.5,2$, $\varepsilon=0.82$, and 
$L_{\rm min}=12$, we obtain 
$z=2.354(3)$, $z=2.356(7)$ and $z=2.358(13)$.
One can also estimate $z$ by fitting 
$z_{\rm eff}(L)$ to $ z + e_{21} L^{-\omega_2}$.  If we determine 
$z_{\rm eff}(L)$ from $\tau_{x=1}(L)$, we obtain
$z=2.357(4)$ for $L_{\rm min}=12$ and $z=2.357(6)$ for $L_{\rm
min}=16$.
All results are perfectly consistent. 
From these analyses we obtain the estimate
\begin{equation}
z=2.356(6)[3],
\label{prelest}
\end{equation}
which is the result of the fit of $\tau_{x=1}(L)$ with
$\varepsilon=0.82$ and $L_{\rm min}=16$. The error in brackets 
gives the variation of the estimate as $\omega_2$ varies within one error bar.

In the above-reported determination we have implicitly
assumed that the RSIM at $p = 0.8$ is exactly improved so that there
are no leading scaling corrections. However, $p^*$ is only known
approximately and thus some residual $O(L^{-\omega})$ scaling
correction are still present.  To determine their relevance, we
exploit the fact that ratios of leading scaling-correction amplitudes 
are universal
and use the bound~\cite{HPPV-07}
\begin{equation}
 |c_{{\cal O} ,11}(p=0.8)/c_{{\cal O} ,11}(p=0.65)| \lesssim 1/30, 
\label{bound-c11}
\end{equation}
which holds for any quantity ${\cal O}$, be it static or dynamic, 
computed in the RSIM at $p=0.8$ and $p = 0.65$ ($c_{{\cal O} ,11}$
is the amplitude of the $L^{-\omega}$ correction appearing in the 
large-$L$ behaviour of ${\cal O}$).
Bound (\ref{bound-c11}) shows that 
$\tau_x(p=0.8;L)^{1+k} \tau_x(p=0.65;L)^{-k}$ is exactly improved 
(the leading correction proportional to $L^{-\omega}$ exactly cancels)
for some $k$ satisfying $|k|\le 1/30$. Thus, an upper bound on the 
systematic error due to the $L^{-\omega}$ 
scaling corrections is obtained by analyzing 
\begin{equation}
\tau_x(p=0.8;L)^{1\pm 1/30} \times \tau_x(p=0.65;L)^{\mp 1/30},
\end{equation}
instead of $\tau_x(p=0.8;L)$. The estimate of $z$ varies by
$\pm 0.008$, which represents the systematic error due to the 
residual $L^{-\omega}$ corrections. The final result is therefore
\begin{equation}
z = 2.356(6)[3]\{8\}.
\label{estimate1}
\end{equation}

\begin{figure*}[tb]
\centerline{\psfig{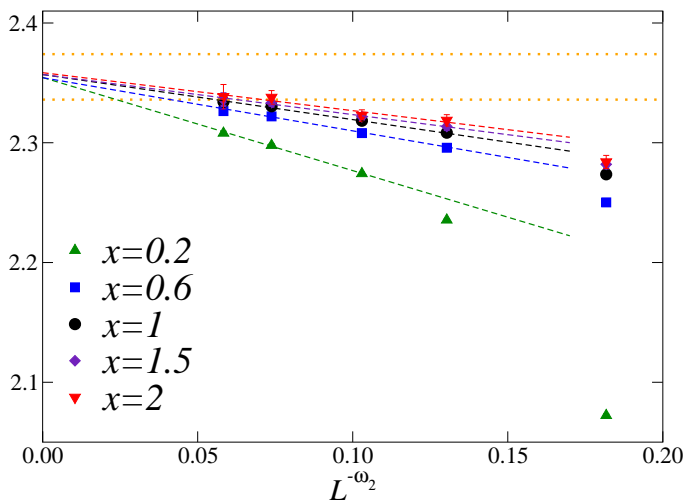}}
\vspace{2mm}
\caption{
Estimates of $z_{\rm eff}(L)$, as obtained from $\tau_x(L)$ 
for several values of
$x$, for the RSIM at $p=0.8$. 
The dotted lines correspond to the result $z = 2.355(16)$,
see (\ref{finalestzim}).
}
\label{taup0.8_x}
\end{figure*}

The above-presented analysis shows that the estimates of $z$ obtained by using
$\tau_x$ with different values of $x$ are perfectly consistent, as of course
should be expected.\footnote{The consistency of the estimates shows also that 
the potential problems due to the Griffiths tail do not occur at the 
values of $x$ and $L$ we consider here.}
There is therefore little advantage in considering many
values of $x$ and it is simpler to restrict the analyses to a single $x$. We
wish to choose it in such a way to minimize scaling corrections and
statistical errors. As is clear from Table~\ref{tabrsimp0.8} statistical
errors decrease with decreasing $x$.  In Fig.~\ref{taup0.8_x} we show $z_{\rm
  eff}(L)$ as computed from $\tau_x$ for different values of $x$. Scaling
corrections decrease with increasing $x$ and are essentially independent of
$x$ for $x \gtrsim 1$. Thus, a good compromise between small statistical
errors and small scaling corrections is obtained by taking $x$ neither too
small nor too large. We have thus chosen $x = 1$.  The quantities that are
analysed in the following sections are always obtained from $\tau_{x=1}(L)$.

\subsubsection{Analysis for $p=0.65$ and $p=0.85$.}
Let us now consider the RSIM at $p=0.65$ and $p=0.85$. Since the model is 
not improved we must include corrections with exponent $\omega$ and 
$2\omega$ at least, i.e. consider correction-to-scaling terms
proportional to $L^{-\omega}\approx L^{-0.29}$ and 
$L^{-2 \omega}\approx L^{-0.58}$, which decrease slower than the 
leading correction term $L^{-\omega_2} = L^{-0.82}$
occurring in improved models.  Assuming this type of corrections,
we fitted $\tau_{x=1}(L)$ with the ans\"atze 
\begin{equation}
\label{ansatz1}
\tau_{x=1}(L) = c L^z \;\;,
\end{equation}
\begin{equation}
\label{ansatz2}
\tau_{x=1}(L) = c L^z \; (1+ c_{11} L^{-\omega}) 
\end{equation}
and 
\begin{equation}
\label{ansatz3}
\tau_{x=1}(L) = c L^z \; (1+ c_{11} L^{-\omega} + c_{12} L^{-2 \omega}) \;\;,
\end{equation}
fixing $\omega=0.29$.  

Let us first discuss the case $p=0.65$. Fits to
(\ref{ansatz1}) give $\chi^2/{\rm DOF} \approx 1$ 
(DOF is the number of degrees of freedom of the fit) starting from
$L_{\rm min}=32$. For $L_{\rm min}=32$ we obtain $z=2.671(5)$. Fits to
(\ref{ansatz2}) give $\chi^2/{\rm DOF} \approx 1$ starting 
from $L_{\rm min}=24$. For $L_{\rm min}=24$ we obtain $z=2.46(2)$.
Fits to (\ref{ansatz3}) 
have $\chi^2/$DOF close to one already for $L_{\rm min}=12$. 
For $L_{\rm min}=12$ and $L_{\rm min}=16$ we obtain
$z=2.31(3)$ and $z=2.23(6)$, respectively.

The same analysis can be repeated for $p=0.85$. If we consider the 
smallest $L_{\rm min}$ corresponding to $\chi^2/$DOF close to 1 for each
fit ansatz, we obtain
$z=2.229(1)$ (fit to (\ref{ansatz1}), $L_{\rm min}=24$),
$z=2.33(3)$  (fit to (\ref{ansatz2}), $L_{\rm min}=12$), and
$z=2.12(2)$  (fit to (\ref{ansatz3}), $L_{\rm min}=8$).
Again, the results of fits to (\ref{ansatz3}) vary significantly with 
$L_{\rm min}$: for $L_{min}=12$ we obtain $z=2.21(8)$.

Using the simple power-law ansatz~(\ref{ansatz1}) one obtains results that 
apparently indicate non-universal, $p$-dependent values of $z$. Including 
the expected corrections to scaling the results for the critical 
dynamic exponent $z$ change rather dramatically, indicating that 
scaling corrections play a crucial role in the analysis. 
However, since the results obtained for $z$ depend strongly on the number of 
correction terms included in the fit and also on the minimal lattice size
$L_{\rm min}$, we cannot obtain a direct accurate estimate of $z$ at these 
values of $p$. Analogously, it is not possible to include 
the additional correction term
$c_{21} L^{-\omega_2}\sim L^{-0.82}$, 
which was important for the analysis at $p=0.8$ (in this case we should also
consider the equivalent correction $L^{-3\omega} \sim L^{-0.87}$).
For these reasons, we do not quote a final result
for $z$ at $p=0.65$ and $0.85$.

\subsubsection{Correction-to-scaling amplitudes.}
\label{univdil}

\begin{figure*}[tb]
\centerline{\psfig{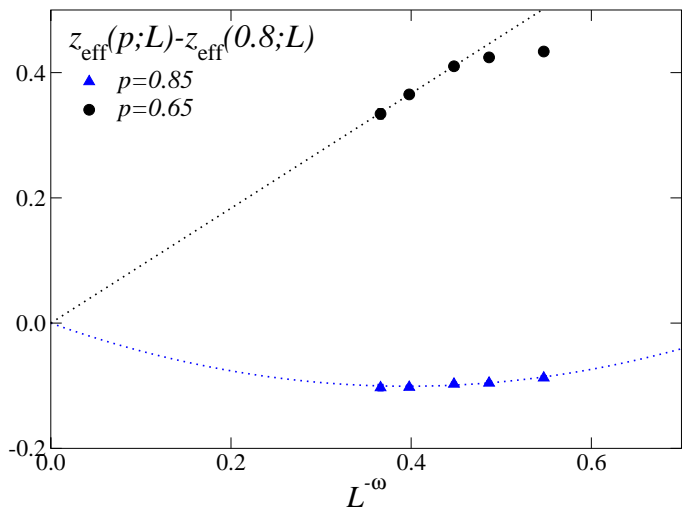}}
\vspace{2mm}
\caption{
Difference 
$\Delta z (p;L) \equiv z_{\rm eff}(p;L)-z_{\rm eff}(p=0.8;L)$ versus
$L^{-\omega}$ with $\omega=0.29$ for the RSIM at $p=0.85,0.65$.  
The dotted lines are the results of
fits to $aL^{-\omega}$ in the case of $p=0.65$, and to
$aL^{-\omega}+bL^{-2\omega}$ for $p=0.85$.
}
\label{zeffd}
\end{figure*}

Here we assume that the value of $z$ is 
universal, i.e. that it does not depend on $p$. Based on this assumption, we
compute amplitude ratios that involve the correction amplitude 
$e_{11}$ defined in (\ref{zefffss}) and verify that these ratios do not depend
on the chosen value of $p$. This provides a consistency check that the 
dynamic universality class is independent of $p$.
This type of analysis
is equivalent in spirit to an analysis in which
data at different values of $p$ are fitted together assuming 
the same dynamic exponent $z$. For instance, this is what was done 
in \cite{BFMMPR-98}. There are, however, two significant differences: 
first, we use the static correction-to-scaling exponents (this allows
us to consider the leading and the subleading scaling correction);
second, we verify 
that the amplitudes of the leading scaling correction satisfy
the constraints imposed by the RG, i.e. we verify the universality
of the amplitude ratios.

For this purpose we consider
\begin{equation}
\Delta z(p;L) \equiv z_{\rm eff}(p;L)-z_{\rm eff}(p=0.8;L).
\end{equation}
For $L\to \infty$ it behaves as 
\begin{equation}
z_{\rm eff}(p;L)-z_{\rm eff}(p=0.8;L)\approx
\bar{e}_{11} L^{-\omega} + \bar{e}_{12} L^{-2\omega} 
+ \cdots +
\bar{e}_{21} L^{-\omega_2} + \cdots 
\label{diffzeff}
\end{equation}
if the dynamic critical behaviour does not depend on $p$.
Since the RSIM at $p = 0.8$ is approximately improved, 
we have $e_{11}\approx 0$ for $p=0.8$, so that 
\begin{equation}
 \bar{e}_{11}\approx e_{11}(p). 
\label{eb11}
\end{equation}
In Fig.~\ref{zeffd} we show the difference (\ref{diffzeff}) as 
obtained from the available data.  Fits of $\Delta z(p;L)$ 
to $aL^{-\omega}$ and $aL^{-\omega}+bL^{-\varepsilon}$
with $\varepsilon=2\omega,\omega_2$ provide estimates of
$e_{11}$.  We obtain $e_{11}=0.9(2)$ at $p=0.65$ and
$e_{11}=-0.55(15)$ at $p=0.85$. As expected, corrections have 
opposite sign in the two cases and are quite significant at the 
present values of $L$. Note that at $p=0.85$ only fits with 
two corrections give a reasonable $\chi^2$, indicating that at least
two correction terms must be taken into account. 

\begin{figure*}[tb]
\centerline{\psfig{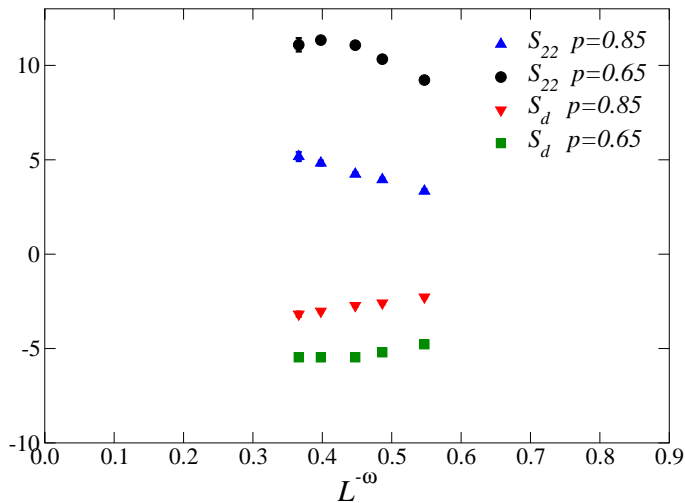}}
\vspace{2mm}
\caption{
Ratios $S_{22}(p;L)$ and $S_d(p;L)$ defined in (\ref{sdef})
for $p=0.85,0.65$, versus $L^{-\omega}$, $\omega = 0.29$.
}
\label{unamrars}
\end{figure*}

Then, we consider the static quartic cumulants 
\begin{eqnarray}
U_{22} \equiv  {\overline{\mu_2^2}-\overline{\mu_2}^2 \over \overline{\mu_2}^2},
\qquad
U_{4}  \equiv { \overline{\mu_4}\over \overline{\mu_2}^{2}}, 
\qquad 
U_d \equiv  U_4-U_{22},
\label{cumulants}
\end{eqnarray}
where $\mu_{k} \equiv \langle \; ( \sum_x \sigma_x\; )^k \rangle$,
at fixed $\xi/L=0.5943$---we call them $\bar{U}_{22}$, $\bar{U}_{4}$, and
$\bar{U}_d$, respectively.
For $L\to\infty$ they behave as
\begin{equation}
\bar{U}_\# = \bar{U}_\#^* + c_{\#,11} L^{-\omega} + \ldots,
\label{u22exp}
\end{equation}
where~\cite{HPPV-07} $\bar{U}_{22}^*=0.148(1)$,
$\bar{U}_{4}^*=1.648(3)$, and $\bar{U}_{d}^*=1.500(1)$.
The ratios of the leading scaling-correction amplitudes are universal. In
the case of $\bar{U}_{22}$ and $\bar{U}_d$, we have~\cite{HPPV-07}
\begin{equation}
s_c={c_{22,11}\over c_{d,11}}=-0.44(3).
\label{cumrac}
\end{equation}
Analogously, the ratio
\begin{equation}
s_\# \equiv {e_{11}\over c_{\#,11}}
\label{cumra}
\end{equation}
is expected to be universal if the dynamic universality class 
is independent of 
$p$. The ratios (\ref{cumra}) can be directly estimated by considering 
\begin{equation}
S_\#(p;L) \equiv {z_{\rm eff}(p;L)-z_{\rm eff}(p=0.8;L)\over
\bar{U}_\#(p;2L) -\bar{U}_\#(p=0.8;2L) }
= 2^\omega s_\# + b_1 L^{-\omega} + b_2 L^{-\omega_2+\omega} + \cdots
\label{sdef}
\end{equation}
In Fig.~\ref{unamrars} we show $S_{22}$ and $S_d$ for $p=0.85,0.65$.  
At $p = 0.65$ a fit of the data with $L\ge 16$ to 
$a+bL^{-\omega}$  ($\omega=0.29$) 
gives $s_{22}=9(1)$ and $s_d=-4.5(5)$.  At $p=0.85$,
the same fit gives
$s_{22}=7.4(9)$ and $s_d=-4.2(5)$.  The agreement is
satisfactory, taking also into account that the errors do not take 
into account several sources of systematic uncertainty.
The approximate $p$-independence 
of the ratios $s_{22}$ and $s_d$ represents a nontrivial
check that the dynamic universality class is independent of $p$.
Assuming universality, we obtain for the RSIM
\begin{equation}
s_d=-4.5(5), \qquad s_{22}=9(2).
\label{unratio}
\end{equation}
Note that these ratios are consistent with 
$s_{d}/s_{22} = s_c =-0.44(3)$, cf. (\ref{cumrac}).

It is interesting to note that the scaling corrections occurring in 
$\tau_x(L)$ 
are significantly larger than those occurring in static quantities.
For instance, we have
\begin{equation}
{c_{11} \over (c_{\#,11}/\bar{U}_\#^*) } = 
     {s_\# U_\#^* \ln 2\over 2^{-\omega} - 1} \approx
     \cases{-5(1) & for $\bar{U}_{22}$ \cr
            26(3)  & for $\bar{U}_{d}$, 
           }
\end{equation}
where $c_{11}$ is defined in (\ref{taufss2}). 

\subsubsection{Improved estimators and universality.}
\label{imprestim}

In the estimate of $z$ obtained at $p=0.8$ in Sec.~\ref{resultsrsimp0.8},
cf.~(\ref{estimate1}), the residual $O(L^{-\omega})$ scaling corrections are
an important source of error.  These corrections can be significantly reduced
by considering improved estimators \cite{HPPV-07}. The estimate of the
universal ratio $s_d$ obtained in Sec.~\ref{univdil} allows us to define
improved quantities with smaller $L^{-\omega}$ scaling corrections. Let us
consider the quantities
\begin{eqnarray}
&& Z_1(r;L)\equiv z_{\rm eff}(L)\left( \bar{U}_d(2L)/\bar{U}_d^*\right)^r,
\qquad 
\nonumber \\
&& Z_2(q;L)\equiv z_{\rm eff}(L) + q (\bar{U}_d(2L) - \bar{U}_d^*).
\end{eqnarray}
For $L\to \infty$ they behave as 
\begin{equation}
Z_i(L)= z + f_{11} L^{-\omega} + f_{12} L^{-2\omega} + \cdots +
f_{21} L^{-\omega_2} + \cdots .
\label{zefffssim}
\end{equation}
where the correction-to-scaling amplitudes depend on $r$ or $q$.
Then, we determine $r^*$ and $q^*$ such that $f_{11}(r^*)=f_{11}(q^*)=0$.
An easy calculation gives
\begin{eqnarray}
r^*= -{ 2^\omega \bar{U}_d^* s_d\over z}, 
\qquad q^*= -2^\omega s_d .
\label{rstar-qstar}
\end{eqnarray}
Note that $r^*$ and $q^*$ are expressed in terms of universal quantities
and thus $Z_1(r^*;L)$ and $Z_2(q^*;L)$ are improved in any model
in the same dynamic universality class. Using $s_d=-4.5(5)$, 
$z = 2.36(2)$, $\bar{U}_d^* = 1.500(1)$, we obtain 
\begin{eqnarray}
r^*= 3.5(5), 
\qquad q^* = 5.5(7).
\label{qstar} 
\end{eqnarray}
The error is mostly due to the error on $s_d$.
For $r=r^*$ and $q=q^*$
the scaling corrections are proportional to 
$L^{-2\omega}$ with $2\omega=0.58(4)$.
In the following, we define improved estimators by taking
the central values of the estimates (\ref{qstar}):
\begin{equation}
z_{\rm im}(L)\equiv Z_1(r=3.5;L),\qquad 
z_{\rm im2}(L)\equiv Z_2(q=5.5;L).
\label{zimdef}
\end{equation}
One may define analogous improved operators by using $\bar{U}_{22}$ instead of
$\bar{U}_d$. Those defined in terms of 
$\bar{U}_d$ are more convenient because $\bar{U}_d^*$
is known with better numerical precision.  
Since $r^*$ and $q^*$
are known only approximately, $z_{\rm im}(L)$ and $z_{\rm im2}(L)$ 
still have $L^{-\omega}$ corrections.
Taking into account the uncertainty on
the estimates of $r^*$ and $q^*$, we obtain the bound
\begin{equation}
|f_{11}/e_{11}|\lesssim 1/6,
\label{f11bound}
\end{equation}
i.e. the 
leading scaling correction in $z_{\rm im}(L)$ and $z_{\rm im2}(L)$ 
is at least a factor of 6 smaller than that occurring in $z_{\rm eff}(L)$.

\begin{figure*}[tb]
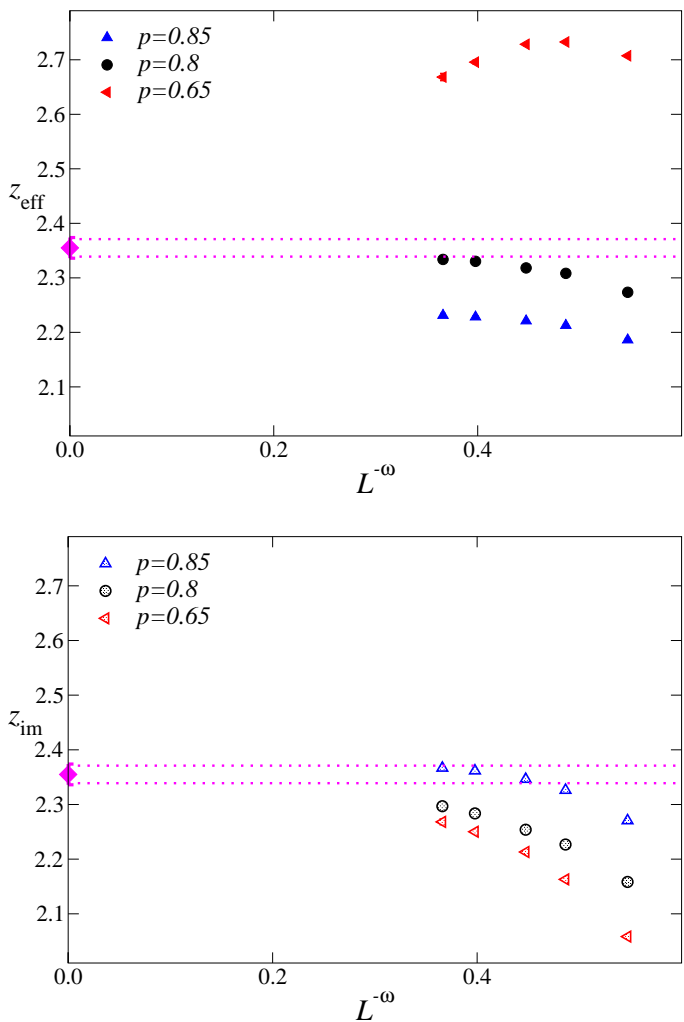

\centerline{\psfig{width=9truecm,angle=0,file=zeffrsx1all.eps}}
\vspace{5mm}
\centerline{\psfig{width=9truecm,angle=0,file=zeffrsx1allim.eps}}
\vspace{2mm}
\caption{
Effective exponents $z_{\rm eff}(L)$ (above)
and $z_{\rm im}(L)$ (below)
versus
$L^{-\omega}$, $\omega=0.29$, for the RSIM at $p=0.65,0.8,0.85$.  
The dotted lines correspond to the final result $z = 2.355(16)$.
}
\label{zrsim-all}
\end{figure*}

In Fig.~\ref{zeffis} we show estimates of $z_{\rm im}(L)$ and $z_{\rm im2}(L)$
at $p=0.8$. Fits to $z+c L^{-\omega_2}$ give results in perfect agreement with
those obtained by fitting $\tau_x(L)$ and $z_{\rm eff}(L)$, see
Fig.~\ref{tausp0.8}.  Note that the data for $z_{\rm im}(L)$ and $z_{\rm
  im2}(L)$ are very close and provide almost equal results. This can be easily
explained by noting that
\begin{equation}
z_{\rm im}(L) - z_{\rm im2}(L) = 
   \left(r - q U_d^*/z\right) b L^{-\omega_2} + \cdots,
\end{equation}
where $b$ is a $p$-dependent coefficient. Since 
$r$ and $q$ are good approximations of $r^*$ and $q^*$ defined in
(\ref{rstar-qstar}), the prefactor is very small, explaining 
why the two quantities behave identically. 
It is interesting to note that $z_{\rm im}(L)$  has
$L^{-\omega_2}$ corrections which are larger than those occurring
in $z_{\rm eff}(L)$, see Fig.~\ref{zeffis}: improved quantities have smaller 
leading scaling corrections but larger subleading ones.

In the following we only 
report results for $z_{\rm im}(L)$.  Fits of $z_{\rm im}(L)$ at $p=0.8$ give
$z=2.355(8)$ [$z=2.356(7)$] for $L_{\rm min}=16$ [$L_{\rm min}=12$], where the
errors also take into account the uncertainty on $\bar{U}_d^*=1.500(1)$.  These
results vary approximately by $\pm 0.008$ when changing $\omega_2$ within
$[0.74,0.90]$, corresponding to the uncertainty on $\omega_2$.
We finally obtain the estimate 
\begin{equation}
z = 2.355(8)[8].
\label{finalestzim}
\end{equation}
Because of the bound (\ref{f11bound}),
the error due to the residual $L^{-\omega}$ scaling corrections is 
negligible. This result confirms the one given in (\ref{estimate1}).  

\begin{figure*}[tb]
\centerline{\psfig{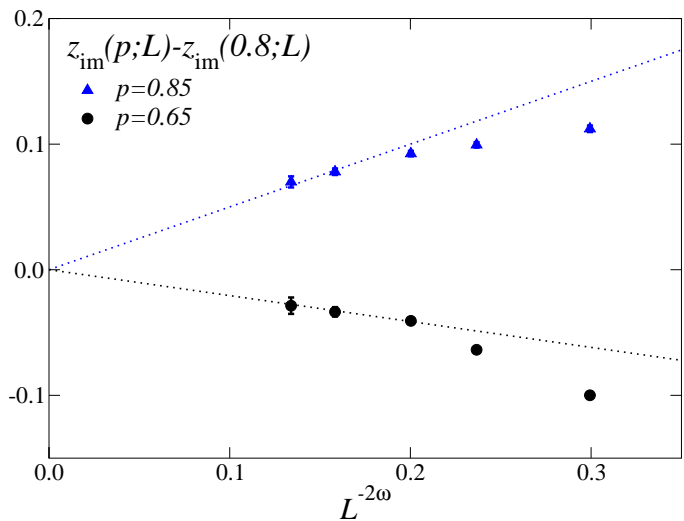}}
\vspace{2mm}
\caption{
Difference $\Delta z_{\rm im} (p;L) \equiv
z_{\rm im}(p;L)-z_{\rm im}(p=0.8;L)$ versus
$L^{-2 \omega}$ with $\omega=0.29$ for the RSIM at $p=0.85,0.65$.  
The dotted lines correspond to fits to $aL^{-2\omega}$.
}
\label{zimrsd}
\end{figure*}

In Fig.~\ref{zrsim-all} we report $z_{\rm im} (L)$ and $z_{\rm eff} (L)$ for
the three values of $p$ we have considered.  In all cases, the improved
exponents are quite close to the final estimate (\ref{finalestzim}). For
$p=0.85$, while $z_{\rm eff}(L)$ was close to 2.25, $z_{\rm im} (L)$ is fully
consistent with 2.355. As for $p=0.65$, the difference between $z_{\rm im}
(L)$ and 2.355 is three times smaller than that between $z_{\rm eff}(L)$ and
2.355.  The still existing discrepancies can be explained by the
next-to-leading $O(L^{-2\omega})$
scaling corrections, as shown by Fig.~\ref{zimrsd} where the difference $\Delta z_{\rm
  im}(p;L) \equiv z_{\rm im}(p;L)-z_{\rm im}(p=0.8;L)$ is plotted versus
$L^{-2\omega}$. Clearly, $\Delta z_{\rm im}(p;L) $ is consistent with zero,
for both $p=0.85$ and $0.65$, if we only consider data with $L\ge 16$,
supporting universality.

In conclusion, the results of the RSIM provide an accurate estimate of
the dynamic exponent $z$, i.e. $z=2.355(16)$, and a robust evidence of
universality, i.e. independence on $p$. 

Finally, we note that the leading amplitude $c$ defined in
(\ref{taufss2}) significantly increases with decreasing
$p$. Indeed, we find $c\approx 0.03,0.05,0.3$ for $x = 1$ and 
$p=0.85,0.8,0.65$, respectively.

\subsection{Universality of $z$ in the RBIM and $\pm J$ Ising model}
\label{univ}

\begin{figure*}[tb]
\centerline{\psfig{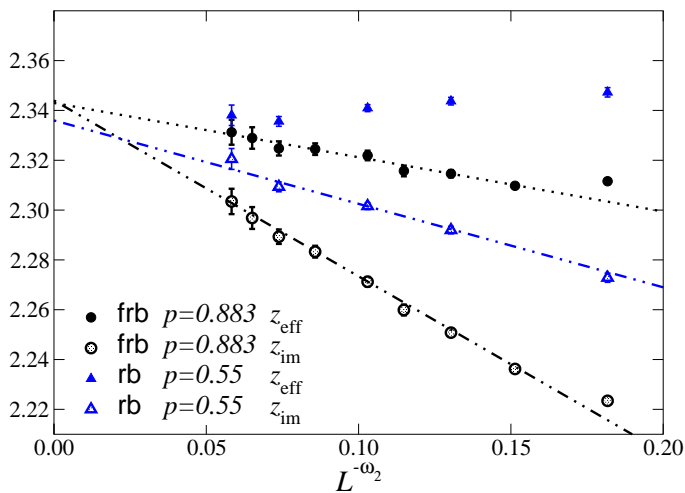}}
\vspace{2mm}
\caption{The effective exponents
$z_{\rm eff}(L)$ and 
$z_{\rm im}(L)$ for the RBIM (rb) at $p=0.55$ and for
the $\pm J$ Ising model (frb) at $p=0.883$,
versus $L^{-\omega_2}$ with $\omega_2=0.82$.
}
\label{zeffibf}
\end{figure*}

We now check the universality of the dynamic exponent $z$ in other RDIs
systems, such as the RBIM and the $\pm J$ Ising model along the
paramagnetic-ferromagnetic transition line.  We first focus on the
approximately improved models, the RBIM at $p = 0.55$ (the RBIM is improved
for $p= p^*=0.54(2)$) and the $\pm J$ Ising model at $p = 0.883$ (improvement
occurs for $p=p^*=0.883(3)$). We perform an analysis analogous to that
presented for the RSIM at $p=0.8$, verifying that scaling corrections decay as
$L^{-\omega_2}$, as expected on the basis of field theory, and computing for
each of them an estimate of $z$. Then, we verify that the results for the
other values of $p$ are consistent with universality, i.e. that the large
observed deviations can be interpreted as scaling corrections.

In Fig.~\ref{zeffibf} we plot $z_{\rm eff}(L)$, defined in (\ref{zeff}), and
$z_{\rm im}(L)$, defined in (\ref{zimdef}), versus $L^{-\omega_2}$ with
$\omega_2=0.82$.  In the case of the $\pm J$ Ising model, both $z_{\rm
  eff}(L)$ and $z_{\rm im}(L)$ clearly show the expected $L^{-\omega_2}$
behaviour. In the case of the RBIM, $z_{\rm im}(L)$ shows a clear linear
trend, while $z_{\rm eff}(L)$ becomes essentially flat as $L$ increases
and is close to the RSIM estimate $z =
2.355(16)$: indeed, $z_{\rm eff}(L) = 2.341(1)$, 2.336(2), 2.338(4) for
$L=16,24,32$.

\begin{figure*}[tb]
\centerline{\psfig{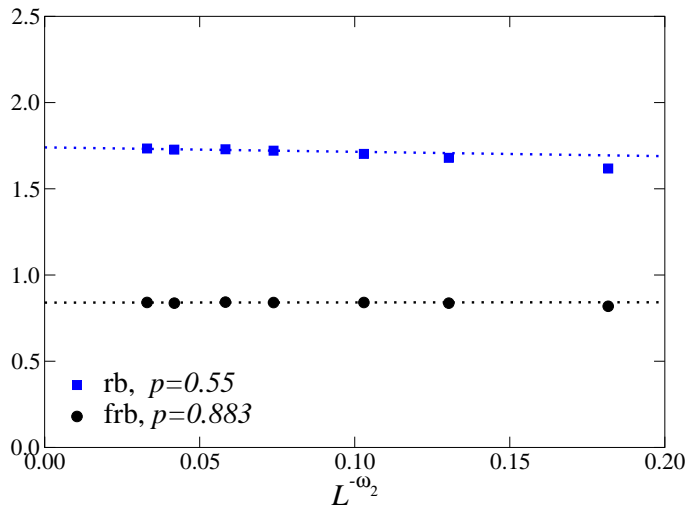}}
\vspace{2mm}
\caption{
The ratio $\tau_{x=1}(p;L)/\tau_{x=1}({\rm RSIM},p=0.8;L)$ for 
the RBIM (denoted by rb) at $p=0.55$ and for the $\pm J$ Ising model (frb)
at $p=0.883$, versus $L^{-\omega_2}$ with $\omega_2=0.82$.  
The dots are the results of fits to $a+cL^{-0.82}$.
}
\label{taubfra}
\end{figure*}

To verify that the RBIM at $p=0.55$ and the $\pm J$ model at $p=0.883$
have the same dynamical critical behaviour as the RSIM at $p=0.8$, 
in Fig.~\ref{taubfra} we plot the ratio
\begin{equation}
\tau_{x=1}(L)/ \tau_{x=1}(L)_{{\rm RSIM}, p=0.8}.
\label{tauratiobf}
\end{equation}
As $L\to \infty$ the data clearly approach a constant, indicating that 
all autocorrelation times diverge with the same $z$.
The data shown in the figure are well fitted to 
$b + b_{2} L^{-\omega_2}$, with $b\approx 1.74$ and $b\approx 0.84$
respectively for the RBIM and $\pm J$ Ising model.

\begin{figure*}[tb]
\centerline{\psfig{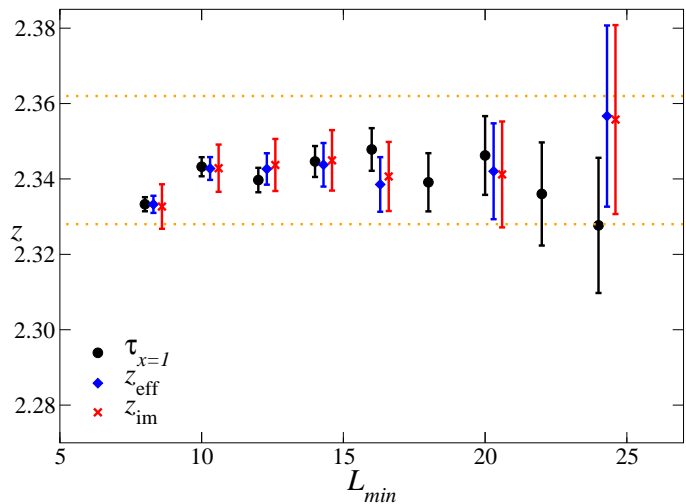}}
\vspace{2mm}
\caption{
Estimates of the dynamic exponent $z$ from fits of $\tau_{x=1}(L)$
to (\ref{acfit}) with $\varepsilon=0.82$, and
of the corresponding  $z_{\rm eff}(L)$ and $z_{\rm im}(L)$ 
to $z+cL^{-0.82}$. The data refer to the 
$\pm J$ Ising model at $p=0.883$. 
Some data are slightly shifted along the $x$
axis to make them visible.  The dotted lines correspond to the final
estimate $z=2.345(17)$, see (\ref{finestpmJ}).
}
\label{taubf0.883}
\end{figure*}

The results of the fits of $\tau_{x=1}(L)$, $z_{\rm eff}(L)$, and $z_{\rm
im}(L)$ for the $\pm J$ Ising model at $p=0.883$ are shown in
Fig.~\ref{taubf0.883}.  In particular, by fitting 
$\tau_{x=1}(L)$ to (\ref{acfit}) with $\varepsilon=0.82$, we obtain
$z=2.345(4)$ and $c=0.0466(8)$ for $L_{\rm min}=14$.
The fit of 
$z_{\rm eff}(L)$ to $z+ e L^{-\omega_2}$ gives $z=2.344(6)$ for 
$L_{\rm min}=14$ and $z=2.342(13)$ for $L_{\rm min}=20$.
These results suggest the estimate $z=2.345(4)[3]$. The error in brackets gives 
the variation of the estimate as $\omega_2$ varies by one error bar.
By fitting 
$z_{\rm im}(L)$ to $z+ e L^{-\omega_2}$ with
$\omega_2=0.82$, we obtain $z=2.345(8)$ (for $L_{\rm min}=14$),
where the error includes the uncertainty on $\bar{U}_d^*$. The
uncertainty on $\omega_2$ changes the estimate by $\pm 0.006$.

As in the case of the RSIM, $p^*$ is only known approximately and thus some
residual leading scaling corrections may still be present.  To determine their
relevance, we again exploit the fact that ratios of amplitudes of leading
scaling corrections are universal, and the bound \cite{HPPV-07-2}
\begin{equation}
 |c_{{\cal O} ,11}(p=0.883)/c_{{\cal O} ,11}(p=0.9)| 
\lesssim 1/5. 
\label{bound-c11-f}
\end{equation}
The error on the estimate obtained from 
$z_{\rm eff}(L)$ due to possible residual $L^{-\omega}$
scaling corrections can be estimated as in the case of the
RSIM at $p=0.8$, obtaining $\pm 0.015$. 
This error is significantly smaller when $z_{\rm im}(L)$ is considered:
the estimate varies by 
$\pm 0.003$.  In conclusion the most precise estimate of $z$ for the 
$\pm J$ Ising model is obtained by using $z_{\rm im}(L)$. We quote
\begin{equation}
z = 2.345(8)[6]\{3\},
\label{finestpmJ}
\end{equation}
which is in good agreement with the RSIM result (\ref{finalestzim}).

Let us now consider the RBIM at $p=0.55$.
By fitting $z_{\rm im}(L)$ 
we obtain $z=2.336(7)$ for $L_{\rm min}=12$, and $z=2.335(9)$ for
$L_{\rm min}=16$.  We obtain
$z=2.335(9)[4]$, where the error in bracket gives the change in the estimate 
as $\omega_2$ varies by one error bar. 
As in the case of the RSIM and the $\pm J$ Ising
model, we must also estimate the error due to the residual
$L^{-\omega}$ scaling corrections.  Using the results
reported in \onlinecite{HPPV-07}, we find that these corrections 
can at most change the estimate of $z$ by $\pm 0.005$.
Our final result is therefore
\begin{equation}
z = 2.335(9)[4]\{5\}\; .
\label{finestrbim}
\end{equation}

\begin{figure*}[tb]
\centerline{\psfig{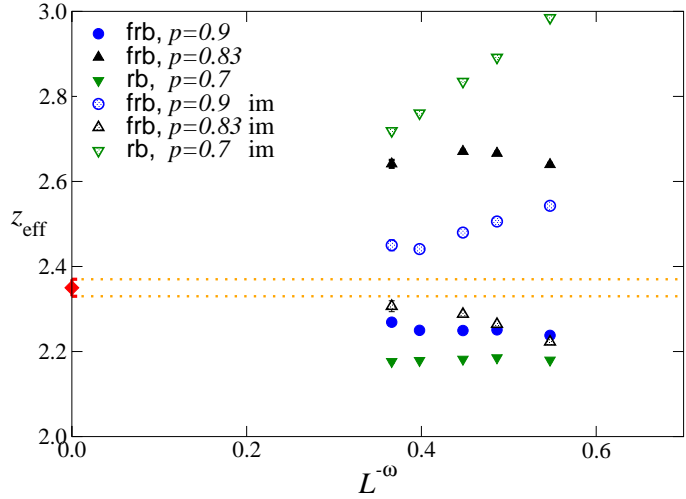}}
\vspace{2mm}
\caption{Effective exponents
$z_{\rm eff}(L)$ and $z_{\rm im}(L)$ 
for the RBIM at $p=0.7$ (rb) and the $\pm J$ Ising model
at $p=0.83,0.90$ (frb).
The dotted lines correspond to the final estimate $z=2.35(2)$. 
}
\label{zeffbf}
\end{figure*}

In Fig.~\ref{zeffbf} we show $z_{\rm eff}(L)$ for other values of $p$,
i.e.~for the RBIM at $p=0.7$ and for the $\pm J$ Ising model at $p=0.83,0.90$.
They are plotted versus $L^{-\omega}$, which is the expected leading
scaling corrections.  As it was observed for the RSIM, 
see Fig.~\ref{figzeffrs},
the results appear strongly $p$-dependent and it is not clear from the 
data that $z_{\rm eff}(L)$ has a model- and $p$-independent limit as
$L\to\infty$. In any case, 
we can show that these data are still consistent with universality if
the expected scaling corrections are taken into account. 
Let us again consider the difference
\begin{equation}
\Delta z_{\rm RS} (p;L) = z_{\rm eff}(p;L)-z_{\rm eff}({\rm RSIM},p=0.8;L),
\label{diffzeff2}
\end{equation}
which, as discussed in the preceding section, should behave as
\begin{equation}
\Delta z_{\rm RS} (p;L) \approx 
\bar{e}_{1}(p) L^{-\omega} + \bar{e}_{2} L^{-\varepsilon},
\label{fitzeff}
\end{equation}
with $\varepsilon=2\omega,\omega_2$ for $L\to\infty$. 
We recall that $\bar{e}_1\approx e_{11}$, cf. (\ref{zefffss}).  
In order to show consistency with
universality, we fit $\Delta z_{\rm RS} (p;L)$ to 
the ansatz (\ref{fitzeff}), determining $e_{11}$. Then, we consider 
$\bar{U}_{22}$ and determine the leading correction-to-scaling 
amplitude $c_{22,11}$ defined in (\ref{u22exp}). Finally,
we verify that the ratio $s_{22}\equiv e_{11}/c_{22,11}$ is independent
of $p$ and of the model, and that it agrees with the 
RSIM estimate (\ref{unratio}), $s_{22} = 9(2)$.

\begin{figure*}[tb]
\centerline{\psfig{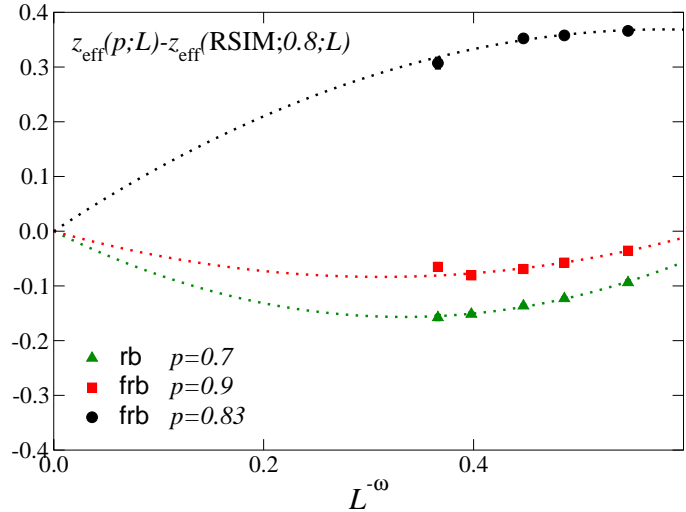}}
\vspace{2mm}
\caption{
The difference $\Delta z_{\rm RS}(p;L) \equiv 
z_{\rm eff}(p;L)-z_{\rm eff}({\rm RSIM},p=0.8;L)$ 
versus $L^{-\omega}$ with $\omega=0.29$ 
for the RBIM (rb) and the $\pm J$ Ising 
model (frb). 
The dotted lines are the results of the
fits to  $aL^{-\omega}+bL^{-2\omega}$.
}
\label{zeffbfd}
\end{figure*}

As shown in Fig.~\ref{zeffbfd}, good fits of $\Delta z_{\rm RS} (p;L)$ to 
(\ref{fitzeff}) are obtained by taking $\varepsilon=2\omega$.
They give $\bar{e}_1 =-1.0(2)$ for the RBIM at
$p=0.7$, and $\bar{e}_1 = 1.2(2),-0.5(1)$ for the $\pm J$ Ising model
at $p=0.83,0.9$, respectively.
The amplitude $c_{22,11}$ can be estimated analogously. We obtain
$c_{22,11}=-0.17(3),\,0.10(2),\,-0.05(1)$,
respectively for the RBIM at $p=0.7$, and the $\pm J$ Ising model
at $p=0.83,0.9$. These results give
\begin{equation} 
s_{22}= 6(2),\;\;12(3),\;\;10(3),
\label{unratio2}
\end{equation}
which are in substantial agreement with the estimate (\ref{unratio}) obtained
from the RSIM. 
These results fully support the interpretation of the anomalous behaviour of the
data shown in Figs.~\ref{zeffbf} and \ref{zeffbfd} as an effect of 
scaling corrections.

\begin{figure*}[tb]
\centerline{\psfig{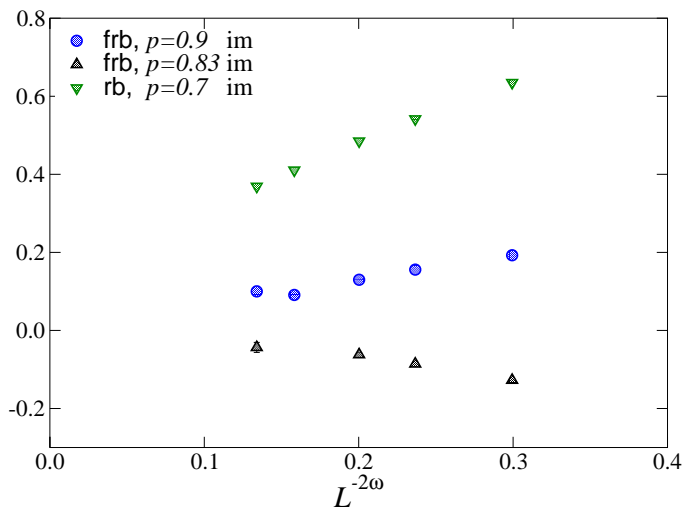}}
\vspace{2mm}
\caption{
The difference $ 
z_{\rm im}(p;L)-2.35$ 
versus $L^{-2\omega}$ with $\omega=0.29$ 
for the RBIM (rb) and the $\pm J$ Ising model (frb). 
}
\label{zimvs2om}
\end{figure*}

In Fig.~\ref{zeffbf} we also show $z_{\rm im}(L)$. For the $\pm J$ model
at $p=0.83$, the improved estimator is significantly 
closer to $z\approx 2.35$ than $z_{\rm eff}(L)$. In the two other cases
deviations are still large, though the data seem to approach faster
the limiting value $z\approx 2.35$ obtained by considering the 
improved models.
In Fig.~\ref{zimvs2om} we show the difference $z_{\rm im}(L) - 2.35$ 
versus $L^{-2\omega}$, which
would be the leading scaling correction if $z_{\rm im}(L)$ were
exactly improved. The data for the $\pm J$ Ising model converge to zero, 
confirming universality. On the other hand, the RBIM data apparently 
extrapolate to a slightly positive value. However, if we include an 
additional correction term (either an $L^{-\omega}$ term, since improvement
is only approximate, or an $L^{-\omega_2}$ term), the data are again
perfectly consistent with universality.

\subsection{Summary of the results for the equilibrium relaxational dynamics at $T_c$}
\label{summequtc}

In this section we have studied the Metropolis equilibrium dynamics in 
the RSIM, the RBIM, and the $\pm J$ Ising model along the
paramagnetic-ferromagnetic transition line, which belong to the 
same static universality class. 
We have verified that the exponent $z$ is the same for the 
RSIM, the RBIM, and the $\pm J$ Ising model for values of the disorder 
parameter $p$ that make these models approximately improved. 
We have obtained $z=2.355(16)$, $z=2.335(18)$, and $z=2.345(17)$
respectively for the RSIM at $p=0.8$, the RBIM at $p=0.55$,
and the $\pm J$ Ising model at $p=0.883$. For the other values of $p$
we have not been able to determine $z$ with comparable precision. We have 
however verified that the dynamic behaviour is always consistent with 
universality once the expected scaling corrections are taken into account. 
In the analyses we have presented, scaling corrections play a very important 
role. We have explicitly verified the FT prediction
that dynamics does not introduce new RG irrelevant operators that are 
more relevant than the two leading ones occurring in the statics. Therefore,
scaling corrections are characterized
by the same universal exponents that control the 
nonasymptotic behaviour in static quantities,
i.e. $\omega=0.29(2)$ and $\omega_2=0.82(8)$. 
As a consequence, the leading $L^{-k\omega}$ scaling corrections 
are absent in dynamic observables 
at the same value $p^*$ of the disorder
parameter $p$ determined by considering static quantities.

Once universality has been checked, we can use our results for the 
RSIM, the RBIM, and the $\pm J$ Ising model, to obtain a final estimate for
$z$. We quote
\begin{equation}
z=2.35(2),
\label{veryfinal}
\end{equation}
which includes all results obtained in the previous sections.

Our result (\ref{veryfinal}) significantly improves earlier MC
estimates~\cite{PV-93,Heuer-93,IIBH-06} of $z$ obtained in equilibrium
MC simulations.
Reference \onlinecite{Heuer-93} considered several values of
$p$ in the range $1>p\ge 0.6$. The final estimate $z=2.4(1)$ 
was essentially derived from the data at $p= 0.8$, where the
finite-size behaviour appeared least dependent on $L$. The
different values of $z$ obtained for other values of $p$ were
interpreted as a crossover phenomenon.  A smaller estimate
$z=2.2(1)$ was found in \onlinecite{IIBH-06}, by MC simulations at
$p=0.85$.  This may be explained by the effect of scaling corrections,
which, as shown by Fig.~\ref{figzeffrs}, give rise to an apparent
smaller value of $z$ if they are not taken into account.

\section{Off-equilibrium estimate of the dynamic critical exponent $z$}
\label{offeq}

The exponent $z$ can also be determined by performing off-equilibrium 
simulations \cite{JSS-89,OJ-95,CG-05}. One starts from a disordered
configuration and observes the relaxation at $T_c$ on sufficiently
large lattices.
In the infinite-volume limit the magnetic
susceptibility $\chi$ is expected to increase with the MC time $t$ as
\begin{eqnarray}
\chi(t) = C_0 t^{\rho} \left( 1 + C_{11} t^{-\upsilon_1} + 
C_{12} t^{-2\upsilon_1} + \cdots + C_{21} t^{-\upsilon_2} + \cdots \right),
\label{chitoff}
\end{eqnarray}
where 
\begin{equation}
\rho = {2-\eta \over z}.
\label{rhoest}
\end{equation}
Using the estimate obtained in Sec.~\ref{zexp}, $z=2.35(2)$, 
and~\cite{HPPV-07} $\eta=0.036(1)$, we predict $\rho=
0.836(7)$.  Moreover, according to the FT perturbative analysis
\cite{JSS-89,OJ-95,CG-05}, the leading scaling-correction exponents
should be the same as those that occur in 
equilibrium (static or dynamic) correlation functions.
Therefore, we expect
\begin{equation}
\upsilon_1 = {\omega\over z} =  0.123(9), \qquad
\upsilon_2 = {\omega_2\over z} =  0.35(3),
\label{expest}
\end{equation}
where we have used $\omega=0.29(2)$ and
$\omega_2=0.82(8)$.  Moreover, 
the leading scaling correction proportional to 
$t^{-\upsilon_1}$ (and also all corrections 
of the form $t^{-k\upsilon_1}$) vanishes in improved models.

Equation (\ref{chitoff}) 
is valid only in the infinite-volume limit. For a finite
system of size $L$ we expect 
\begin{equation}
\chi(t,L) = C_0 \,t^\rho \,  E_0(t L^{-z}) (
1 + C_{11} t^{\upsilon_1} E_1(t L^{-z})  + \cdots)
\label{fsschit}
\end{equation}
where $E_i(x)$ are universal functions satisfying $E_i(0)=1$ and 
$E_0(x)\sim x^{-\rho}$, $E_1(x)\sim x^{\upsilon_1}$ for large $x$.

The off-equilibrium relaxational dynamics of the RSIM has already been
investigated in \onlinecite{PRR-99,SP-05} for various values of $p$ in
the range $1>p\ge 0.4$.  Their results do not agree with the
above-reported predictions.  
References \onlinecite{PRR-99,SP-05} obtain $z=2.62(7)$ and $z=2.6(1)$, 
respectively, independently of the dilution
parameter $p$. They also estimate the leading correction-to-scaling 
exponent $\upsilon_1$. Given their estimate of $z$, this allows them to 
estimate $\omega$. They quote
$\omega=0.50(13)$ and $\omega=0.61(6)$, respectively.
It is quite difficult to reconcile these results with the FT predictions;
in particular, the absence of corrections proportional to $t^{-0.12}$, 
i.e. controlled by the leading exponent $\omega = 0.29(2)$,
is quite surprising. Another numerical
MC work~\cite{IOK-99} investigated the nonequilibrium relaxation dynamics of
the $\pm J$ Ising model, and showed an apparent non-universal dynamical
critical behaviour along the paramagnetic-ferromagnetic transition line.
Also these results are in contrast with 
the FT predictions reported at the beginning of section. 

In the following we further investigate this issue.  We study
the Metropolis dynamics of the RSIM at $p=0.8$ after a
quench from $T=\infty$ to $T_c$.  This represents a nontrivial check
of the FT predictions, since the estimates
(\ref{rhoest}) and (\ref{expest}) are quite precise. 
Since the model is approximately improved
[$p^*=0.800(5)$], we expect that
$C_{1k}$ vanishes for all values of $k$, 
and thus we predict
\begin{eqnarray}
&&\chi(t) = C_0 t^{\rho} \left( 1 + C_{21}t^{-\upsilon_2} + \cdots \right),
\label{chitoffi}\\
&&\rho=0.836(7), \qquad \upsilon_2=0.35(3).
\nonumber
\end{eqnarray}
As in the equilibrium case, 
we define an effective exponent 
\begin{equation}
\rho_{\rm eff}(t) \equiv {\ln[\chi(2t)/\chi(t)]\over \ln 2},
\label{rhoeffdef}
\end{equation}
which behaves as 
\begin{equation}
\rho_{\rm eff}(t) = \rho + c t^{-\upsilon_2} + \ldots
\label{ivrhoeff}
\end{equation}
for $t\to \infty$. On a finite lattice, (\ref{ivrhoeff}) is 
replaced by 
\begin{equation}
\rho_{\rm eff}(t,L) = \rho + e_0(t L^{-z})
\label{rhoeffsca}
\end{equation}
where we have neglected large-$t$ (scaling) corrections and 
$e_0(x)$ is a universal function (apart from a normalization of the 
argument) such that $e_0(0) = 0$ and $e_0(x)\to -\rho$ for $x\to\infty$.

\begin{figure*}[tb]
\centerline{\psfig{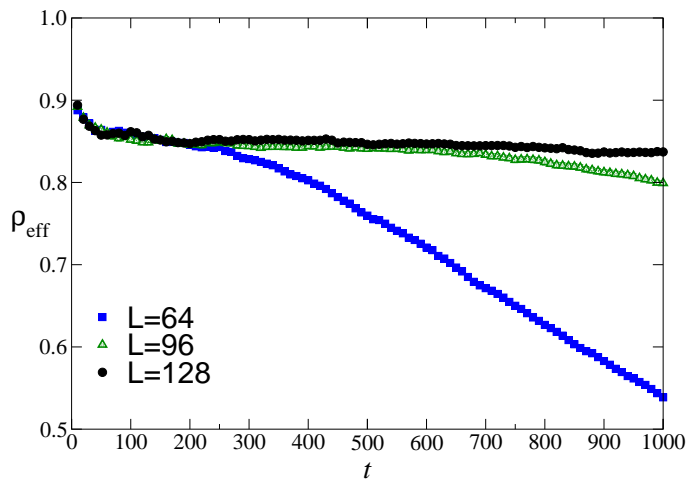}}
\vspace{2mm}
\caption{
Effective exponent $\rho_{\rm eff}(t,L)$ for $L=64,96,128$. Here $t$ is 
the number of MC sweeps.
}
\label{rhoeff}
\end{figure*}

\begin{figure*}[tb]
\centerline{\psfig{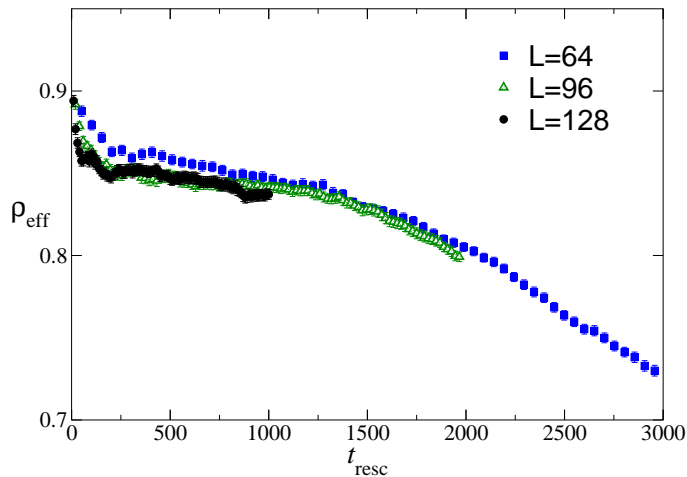}}
\vspace{2mm}
\caption{
The effective exponent
$\rho_{\rm eff}(t,L)$ versus $t_{\rm resc}=t (L/128)^{-z}$.
}
\label{rhoeffsc}
\end{figure*}

We have
performed off-equilibrium MC simulations on lattices of size
$L=64,96,128$ at $\beta = 0.2857430$
[our presently best estimate of $\beta_c$ is $\beta = 0.2857431(3)$].
For each lattice size we average over $N_s = 320000$ disorder
configurations. For each disorder configuration we start from a (different)
$T=\infty$ configuration and perform 2000 Metropolis sweeps, using the
algorithm described in Sec.~\ref{MC} and \ref{appmetro}.  In
Fig.~\ref{rhoeff} we show $\rho_{\rm eff}(t,L)$ for $L=64,96,128$. It 
clearly shows finite-size corrections, and, for each $L$, 
$\rho_{\rm eff}(t,L)$ follows the 
infinite-volume curve up to an $L$-dependent  value $t_{\rm max}(L)$. As
shown by Fig.~\ref{rhoeffsc}, where $\rho_{\rm eff}(t,L)$ is plotted versus 
$t_{\rm resc}\equiv t (L/128)^{-z}$, finite-size effects are
consistent with (\ref{rhoeffsca}).  Thus, the value $t_{\rm max}(L)$,
after which finite-size effects cannot be neglected, increases as $L^{z}$.  
Infinite-volume quantities, such as $\rho_{\rm eff}(t)$, must be
obtained from the data at $t<t_{\rm max}(L)$. Fig.~\ref{rhoeff} indicates that,
with the statistical errors of our data, $t_{\rm max}(L) \approx 120, 600$ for
$L\approx 64, 128$. Since $\rho_{\rm eff}(t,L)$ 
is defined using data at $t$ and
$2t$, this implies that, for $L=128$, only data corresponding to 
$t\lesssim 1200$ have
negligible finite-size effects within our error bars.  Finite-size effects
give rise to a systematic error in the estimate of
$\rho$. As is clear from Fig.~\ref{rhoeff}, 
they yield smaller values of $\rho$, and therefore larger
values of $z$.

\begin{figure*}[tb]
\centerline{\psfig{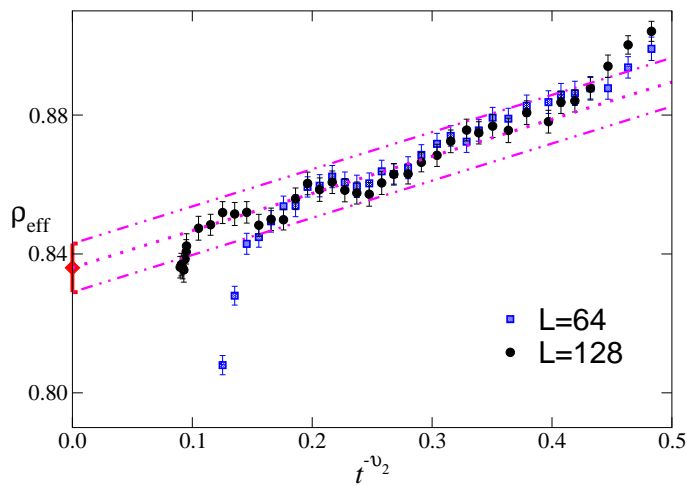}}
\vspace{2mm}
\caption{
Effective exponent $\rho_{\rm eff}(t,L)$ versus $t^{-\upsilon_2}$ with 
$\upsilon_2=0.35$. Finite-size effects are negligible for 
$t^{-\upsilon_2}\gtrsim 0.10$ ($L=128$) and 
$t^{-\upsilon_2}\gtrsim 0.17$ ($L=64$).
The lines are drawn to guide the eyes.
}
\label{rhoeffvst35}
\end{figure*}

\begin{figure*}[tb]
\centerline{\psfig{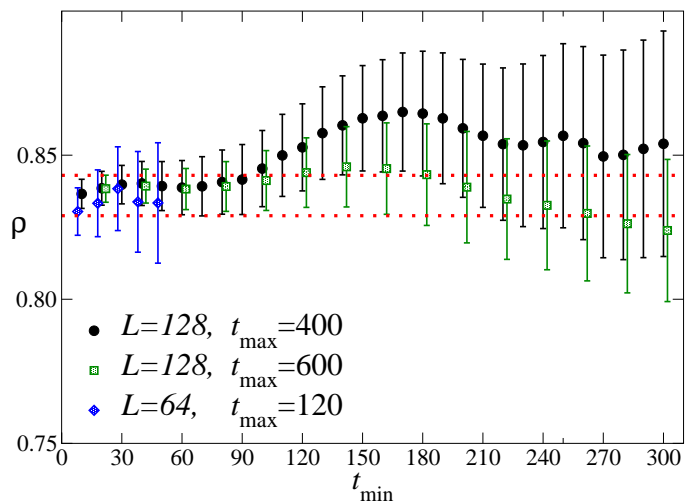}}
\vspace{2mm}
\caption{
Results of fits of $\rho_{\rm eff}(t,L)$ for
$t_{\rm min}\le t \le t_{\rm max}$ to
$\rho+c t^{-\upsilon_2}$ with $\upsilon_2=0.35$,
for $L=64,128$.
The dotted lines correspond to the prediction $\rho=0.836(7)$ obtained
by using the equilibrium result $z = 2.35(2)$ and 
$\eta = 0.036(1)$ \cite{HPPV-07}.
}
\label{fitrho}
\end{figure*}

In Fig.~\ref{rhoeffvst35} we plot $\rho_{\rm eff}(t,L)$ 
for $L=64,128$ versus
$t^{-\upsilon_2}$ with $\upsilon_2=0.35$. 
Finite-size effects are negligible for 
$t^{-\upsilon_2} > t_{\rm max}(L)^{-\upsilon_2}\approx 0.17, 0.10$,
for $L=64,128$, respectively.
The data satisfying this inequality
clearly follow a unique curve, which is expected to
behave as $\rho+ c t^{-\upsilon_2}$ for sufficiently large values of
$t$. The data plotted in Fig.~\ref{rhoeffvst35} clearly show such a
behaviour in the region $t^{-\upsilon_2}\lesssim 0.4$ (corresponding to
$t\gtrsim 10$), and are perfectly compatible with the values
$\rho=0.836(7)$ and $\upsilon_2=0.35(2)$.  This is also shown by
Fig.~\ref{fitrho}, where we plot the results of fits of 
$\rho_{\rm eff}(t,L)$ for $t_{\rm min}\le t \le t_{\rm max}$ to $\rho+c
t^{-\upsilon_2}$ with $\upsilon_2=0.35$, for $L=64,128$.

The above results provide a nice check of the results of the previous section
and confirm the RG prediction
that the off-equilibrium relaxational critical dynamics is characterized by the
same dynamic exponent $z$ and the same 
scaling-correction exponents $\omega$ and
$\omega_2$ as the equilibrium critical dynamics.  Note that the
results of Fig.~\ref{rhoeffvst35} rule out the larger
estimates of $z$ obtained in \onlinecite{PRR-99,SP-05},
$z=2.62(7)$ and $z=2.6(1)$, corresponding to
$\rho=0.750(20)$ and $\rho=0.755(29)$,
respectively (using~\cite{HPPV-07} $\eta=0.036(1)$).

\section{Conclusions}
\label{conclusions}

In this paper we have studied the purely relaxational dynamics (model A) in
randomly diluted Ising (RDIs) systems.  According to standard RG arguments
applied to dynamics, the dynamic critical behaviour in such systems should
belong to the same model-A dynamic universality class.  If this description is
correct, the dynamic exponent $z$ is the same in all RDIs systems and the
leading scaling corrections are controlled by the same RG operators that
appear in the statics and therefore are characterized by the static
correction-to-scaling exponents $\omega=0.29(2)$ and $\omega_2=0.82(8)$.  For
the same reasons, in the case of improved Hamiltonians, leading scaling
corrections should also be absent in dynamical quantities. Therefore, improved
models are expected to provide the most precise estimates of universal dynamic
quantities. For instance, in FSS studies at the critical point, corrections to
scaling decay as $L^{-0.82}$ in improved models, while in generic RDIs systems
the approach to the infinite-volume limit is much slower, corrections decaying
as $L^{-0.29}$.

The main results of our analysis can be summarized as follows.
\begin{itemize}
\item[(i)] We consider the RSIM at $p=0.8$, the RBIM at $p=0.55$, and the $\pm
  J$ Ising model at $p=0.833$, at the critical point. These three models are
  approximately improved (the best estimates of $p^*$, the value of the
  disorder parameter corresponding to an improved model, are 0.800(5),
  0.54(2), 0.833(3) in the three models, respectively
  \cite{HPPV-07,HPPV-07-2}). We perform high-statistics equilibrium MC
  simulations on lattices $L^3$, $L\le 64$, using the Metropolis algorithm
  (for the RSIM and the RBIM a small modification is needed to ensure
  ergodicity, see \ref{appmetro}). We determine the exponent $z$, obtaining $z
  = 2.355(16)$ for the RSIM, $z = 2.335(18)$ for the RBIM, and $z = 2.345(17)$
  for the $\pm J$ Ising model.  These results are in perfect agreement,
  providing strong support to the FT prediction that all RDIs models belong to
  the same dynamic model-A universality class. We also investigate in detail
  the scaling corrections: they are perfectly consistent with a behaviour of
  the form $L^{-\omega_2}$, with $\omega_2 = 0.82(8)$. Again this is in
  agreement with the FT analysis.  Our final result is
\begin{equation}
   z = 2.35(2).
\end{equation}
Note that, while we confirm the general scenario predicted by field theory,
there is a quantitative difference between our result and that 
obtained by resumming the perturbative expansions at two and three loops,
$z\approx 2.18$ \cite{BDFH-05,PPK-06}.
This may be due to a poor convergence of the perturbative
FT expansions. The apparent agreement with the $O(\sqrt{\epsilon})$
result~\cite{GMM-77} $z= 2 + \sqrt{6\epsilon/53}$, which would give
$z\approx 2.336$ for $\epsilon=1$, is likely only by chance.

\item[(ii)]
We investigate the Metropolis dynamics in equilibrium in
the RSIM, the RBIM, and in the $\pm J$ Ising model for other values of 
$p$. Here, as expected, corrections are very strong. In the FSS 
analysis, the leading term is expected to decay as $L^{-\omega}$, 
$\omega = 0.29(2)$. We are not able to determine $z$ in 
these models as accurately as in improved models.
In any case we verify that the $L$-behaviour of the MC data at $T_c$ 
is consistent with universality and with the constraints imposed by the RG 
(universality of ratios of correction-to-scaling amplitudes).

\item[(iii)] We have no evidence of two different universality 
classes depending on the disorder strength
\cite{MKB-04,PPVK-07}. In particular,
we show that the critical behavior 
is not influenced by the geometrical structure of the vacancies and 
does not depend whether the vacancies percolate or not. 
Indeed, since site and bond vacancies 
percolate for $p \lesssim 0.688$ and $p \lesssim 0.751$, respectively,
in the improved RSIM ($p = 0.8$) vacancies form finite clusters, while in the 
improved RBIM ($p = 0.55$) vacancies percolate. 
Nonetheless, the critical behavior is the same.

\item[(iv)]
The results for the $\pm J$ Ising model
show that frustration is irrelevant for the relaxational behaviour along the
paramagnetic-ferromagnetic transition line. It is worth mentioning
that this is not true for the cluster dynamics. In that case 
the exponent $z$ in the $\pm J$ Ising model is much larger than 
in the RSIM and RBIM. In the frustrated case we obtained 
$z\approx 1.6$~\cite{HPPV-07-2}, while in the second one simulations indicate
$z\lesssim 0.5$~\cite{IIBH-06}.

\item[(v)]
We investigate the off-equilibrium relaxational dynamics in the 
RSIM at $p = 0.8$. We start from disordered $T=\infty$ configurations 
and observed the relaxation at $T = T_c$. 
The results show that our equilibrium estimate $z=2.35(2)$
is perfectly consistent with the off-equilibrium MC data.
In the analysis particular care has been taken to avoid finite-size 
corrections. Moreover, the large-time scaling
corrections are consistent with what is expected 
on the basis of field theory \cite{JSS-89,OJ-95,CG-05}. Since the 
model is improved, we do not observe corrections proportional to 
$t^{-\omega/z}$;
instead our data show corrections that are proportional to
$t^{-\omega_2/z}$. Here $\omega$ and $\omega_2$ are the static 
correction-to-scaling exponents, $\omega=0.29(2)$ and $\omega_2=0.82(8)$.
\end{itemize}

The relaxational critical behaviour within the
RDIs universality class is also relevant for the so-called model-C
dynamics, where the order parameter couples with a conserved
secondary density~\cite{HH-77}.  In the case of the 3D RDIs
universality class, the asymptotic critical behaviours of the model-A and 
model-C
dynamics are described by the same stable fixed point. Therefore, they
are expected to have the same dynamic exponent $z$.  This is
essentially related to the fact that the specific-heat exponent of RDIs
systems, $\alpha=-0.049(6)$, is negative~\cite{HHM-74}.  A drastic
change occurs in the approach to the asymptotic behaviour,
which is expected to be much slower in model C
\cite{DFHM-05}. The coupling with a conserved scalar density
gives rise to very slowly decaying $O(\xi^{-\omega_c},L^{-\omega_c})$
scaling corrections with\footnote{
When the specific-heat exponent $\alpha$ is negative, the asymptotic
  critical behaviour of model C is the same as that of model A, because they
  have the same stable fixed point \cite{HHM-74,HH-77}.  Nevertheless,
  the coupling between the order parameter $\varphi(x)$ and the
  conserved scalar density $\varepsilon(x)$ introduces a new irrelevant
  RG perturbation, which is not present in the model A and
  which gives rise to very
  slowly decaying scaling corrections.  The RG dimension $y_c$ of the
  Hamiltonian coupling term 
  ${\cal H}_{\varphi\varepsilon} = \gamma_0 \int d^d x
  \,\varepsilon \varphi^2$ can be computed by using
  nonperturbative scaling arguments: $y_c=y_{\varphi^2} +
  y_\varepsilon - d = 1/\nu-d/2$.  This implies that there are
  $O(\xi^{-\omega_c})$ scaling corrections to the asymptotic critical
  behaviour, with $\omega_c=-y_c$. Using
  \cite{HPPV-07} $\nu=0.683(2)$, one obtains
  $\omega_c=0.036(4)$.} 
\begin{equation}
\omega_c={3\over 2}-{1\over \nu} = -{\alpha\over 2\nu} = 0.036(4),
\label{yc}
\end{equation}
which is much smaller than the leading scaling-correction exponent of the
model-A dynamics, which is $\omega=0.29(2)$.

\section*{Acknowledgments}

We thank Pasquale Calabrese for very useful discussions.  The MC simulations
have been done at the Computer Laboratory of the Physics Department at 
Pisa University.

\appendix

\section{Estimate of the leading correction-to-scaling exponent $\omega$}
\label{omegaest}

In this appendix we compute the leading
correction-to-scaling exponent $\omega$.
We use the method discussed in \onlinecite{HPPV-07}
and combine the data  of
\onlinecite{HPPV-07} with those obtained here.  We consider the
quartic cumulants $\bar{U}_{22}$ and $\bar{U}_d$,
cf. (\ref{cumulants}), at fixed $\xi/L=0.5943$, at
$p=0.85,0.8.0.65$ on lattices of size $L\le 192$.

\begin{figure*}[tb]
\centerline{\psfig{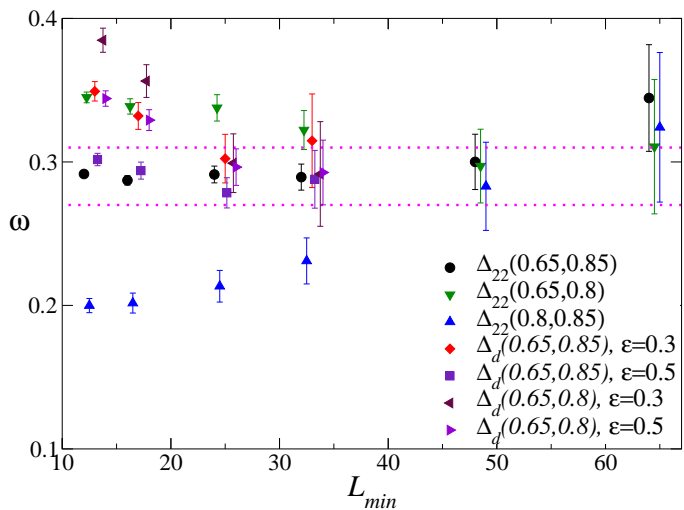}}
\vspace{2mm}
\caption{
Estimates of the leading correction-to-scaling exponent $\omega$.
The dotted lines correspond to our final result $\omega=0.29(2)$.
}
\label{omegafig}
\end{figure*}

As in \onlinecite{HPPV-07}, 
in order to estimate $\omega$, we consider the differences
\begin{eqnarray}
&&\Delta_{22}(p_1,p_2;L) \equiv \bar{U}_{22}(p_1;L) - \bar{U}_{22}(p_2;L) ,
\label{deltau22d}\\
&&\Delta_{d}(p_1,p_2;L) \equiv \bar{U}_{d}(p_1;L) - \bar{U}_{d}(p_2;L).
\end{eqnarray}
Universality implies that
\begin{equation}
\Delta\approx 
c_{\Delta,11} L^{-\omega} + c_{\Delta,12} L^{-2\omega} + \cdots + 
  c_{\Delta,21} L^{-\omega_2} +\cdots
\label{eq-sec5.3}
\end{equation}
In the case of $\Delta_{22}$,
fits to $c L^{-\omega}$ provide good and stable
results. In the case of $\Delta_d$, an additional
correction term is needed in order to
obtain an acceptable $\chi^2$. Hence, we fit $\Delta_d$ to  
$c_1 L^{-\omega} + c_2 L^{-\varepsilon}$.
In Fig.~\ref{omegafig} we show the results as a function of $L_{\rm min}$,
the minimum lattice size used in the fit.
They become independent of $p_1$, $p_2$, and of the considered quantity as 
$L_{\rm min}$ increases.
The most stable results are obtained by taking $p_1=0.85$ and $p_2=0.65$.
From the results shown in Fig.~\ref{omegafig}, we obtain the estimate
\begin{equation}
\omega=0.29(2).
\end{equation}
This estimate is more precise than previous ones and is consistent with 
$\omega=0.33(3)$ obtained in \onlinecite{HPPV-07} by using part of the
data at $p=0.65$ and $p=0.8$.  For comparison, the FT six-loop result
\cite{PV-00} is $\omega=0.25(10)$ (we also mention the five-loop result
$\omega = 0.32(6)$ of \onlinecite{PS-00}).

\section{Metropolis algorithm for randomly diluted Ising models}
\label{appmetro}

We have implemented  the standard local Metropolis algorithm 
with the acceptance rate 
\begin{equation}
\label{accept}
 A = \mbox{min}[1,\exp(-\beta [\cal{H}' - \cal{H}])],
\end{equation}
where $\cal{H}'$ and $\cal{H}$ correspond to 
the Hamiltonian evaluated for the proposal
and for the given spin configuration, respectively. 
The proposal is generated by
flipping the sign of the spin at a single site $x$ of the lattice. Hence
$\cal{H}' - \cal{H}$ depends only on the values of the spins at the site $x$ 
and at its neighbours $y$.

To speed up the simulation we use
multispin coding (see, e.g., \cite{multispin}), evolving in parallel
$n_{\rm bit}$ systems ($n_{\rm bit} = 64$ in our case).
For each of them we use a different set
of disorder variables.  The implementation in the RSIM and RBIM is more
complicated than in the standard Ising model and the $\pm J$ Ising model, 
since the
sum over the nearest neighbours assumes one of the 13 values
$\{-6,-5,...,5,6\}$ and not only the 7 even values $\{-6,-4,...,4,6\}$.

We perform the single-site update sequentially, moving from one site to one of
its neighbours in a typewriter fashion. This causes problems with
ergodicity.
This can be understood by considering the isolated lattice sites, i.e.
the sites $x$ such that $\rho_y=0$ (RSIM)
or $J_{<xy>}=0$ (RBIM) for all neighbours $y$. 
For an isolated site the difference
$\cal{H}'-\cal{H}$ always vanishes, so that,
using the acceptance rate (\ref{accept}), the proposal 
is always accepted.  Hence, going through the lattice twice, the
spins on the isolated sites go back to their values. Therefore, the
configuration restricted to the isolated sites is not changed.
Note that the problem is not restricted to isolated sites only.  For
example, for the one-dimensional chain one can easily prove that a
regular update sweep using the acceptance rate (\ref{accept}) is not
ergodic.  There are many ways to avoid this problem. For performance
reason, we prefer to update the spins sequentially. To avoid the problem
the spin flip is proposed with probability $w$ strictly smaller than
one. We have chosen $w=0.9$. Note that the problem occurs only in the RSIM and 
in the RBIM. For the $\pm J$ Ising 
model the standard Metropolis update can be used.

\end{document}